\documentclass[journal]{IEEEtran}
\usepackage{cite}
\usepackage{graphicx}
\usepackage{amsmath}
\usepackage{times}
\usepackage{multirow}
\usepackage{amssymb}
\usepackage{subfigure}
\usepackage[linesnumbered,ruled,vlined]{algorithm2e}
\usepackage{enumerate}
\usepackage{hyperref}
\usepackage{color}
\usepackage{array}
\usepackage{footnote}
\usepackage{bigstrut}
\usepackage{diagbox}
\usepackage{rotating}
\usepackage{makecell}
\usepackage{comment}

\begin{document}
	\title{TSC-PCAC: Voxel Transformer and Sparse Convolution Based Point Cloud Attribute Compression for 3D Broadcasting}
	\author{Zixi Guo, Yun Zhang,~\IEEEmembership{Senior Member,~IEEE}, Linwei Zhu, Hanli Wang,~\IEEEmembership{Senior Member,~IEEE}, \\
Gangyi Jiang,~\IEEEmembership{Senior Member,~IEEE}
		\thanks{Zixi Guo and Yun Zhang are with the School of Electronics and Communication Engineering, Shenzhen Campus, Sun Yat-Sen University, Shenzhen 518017, China. (Email: guozx29@mail2.sysu.edu.cn, {zhangyun2}@mail.sysu.edu.cn).}
		\thanks{Linwei Zhu is with the Shenzhen Institute of Advanced Technology, Chinese Academy of Sciences, Shenzhen 518055, China (e-mail: lw.zhu@siat.ac.cn).}
		\thanks{Hanli Wang is with the Department of Computer Science and Technology, Tongji University, Shanghai 200092, China (e-mail: hanliwang@tongji.edu.cn).}
\thanks{Gangyi Jiang is with the Faculty of Information and Science and Engineering, Ningbo University, Ningbo 315211, China (e-mail: jianggangyi@nbu.edu.cn).}
	}
	
\markboth{IEEE Transactions on BROADCASTING}%
{Guo \MakeLowercase{\textit{et al.}}: }
	
	
	\maketitle
	\begin{abstract}
			Point cloud has been the mainstream representation for advanced 3D applications, such as virtual reality and augmented reality. However, the massive data amounts of point clouds is one of the most challenging issues for transmission and storage. In this paper, we propose an end-to-end voxel Transformer and Sparse Convolution based Point Cloud Attribute Compression (TSC-PCAC) for 3D broadcasting. Firstly, we present a framework of the TSC-PCAC, which includes Transformer and Sparse Convolutional Module (TSCM) based variational autoencoder and channel context module. Secondly, we propose a two-stage TSCM, where the first stage focuses on modeling local dependencies and feature representations of the point clouds, and the second stage captures global features through spatial and channel pooling encompassing larger receptive fields. This module effectively extracts global and local inter-point relevance to reduce informational redundancy. Thirdly, we design a TSCM based channel context module to exploit inter-channel correlations, which improves the predicted probability distribution of quantized latent representations and thus reduces the bitrate. {Experimental results indicate that the proposed TSC-PCAC method achieves an average of 38.53\%, 21.30\%, and 11.19\% bitrate reductions {on datasets 8iVFB, Owlii, 8iVSLF, Volograms, and MVUB compared to the Sparse-PCAC, NF-PCAC, and G-PCC v23 methods, respectively.}  The encoding/decoding time costs are reduced 97.68\%/98.78\% on average compared to the Sparse-PCAC. The source code and the trained TSC-PCAC models are available at
		\href{https://github.com/igizuxo/TSC-PCAC}{https://github.com/igizuxo/TSC-PCAC.}}
	\end{abstract}
	
	
	\section{Introduction}
	
	\IEEEPARstart{W}{ith} the development of information technology, the demands for more realistic visual entertainment have been rising continuously and rapidly. 3D visual application has emerged as a pivotal trend {nowadays} as it provides 3D depth perception and immersive visual experience. {Point cloud has come into prominence as an indispensable form of expression within the realm of 3D vision, which has wide applications in entertainment, medical imaging, engineering, navigation and autonomous vehicles.} However, a high-quality large-scale point cloud at one instant contains millions of points, where each point consists of 3D geometry and high dimensional attributes, such as color, transparency, reflectance and so on. These massive data {poses} great challenges for transmission and broadcasting, which {hinders} the widespread application of point clouds. Therefore, there is a pressing demand for point cloud compression that can substantially reduce the size of point cloud data. Nevertheless, unlike images where elements are densely and regularly distributed in 2D plane, 3D point cloud is irregular and sparse, leading to irregular 3D patterns and low correlations between neighboring points. These inherent characteristics {pose} substantial challenges to point cloud compression.

	To compress the point cloud effectively, the Moving Picture Expert Group (MPEG) has proposed two traditional point cloud compression methods, i.e.,  Geometry-based Point Cloud Compression (G-PCC) \cite{mammou2019g} and Video-based Point Cloud Compression (V-PCC) \cite{mammou2017video}.
	The G-PCC directly encodes point clouds with an octree structure in 3D space. On the other hand, V-PCC projects the 3D point clouds into 2D images, which are then encoded using conventional codecs in 2D space, such as High Efficiency Video Coding (HEVC) and the latest Versatile Video Coding (VVC).
	{In addition to traditional compression methods, given the remarkable achievements in deep learning-based image compression technologies \cite{13}, many researchers have recently started exploring the potential of deep learning-based point cloud compression techniques including geometry and attribute \cite{huang20193d,5,4,15,16,wang2023lossless,nguyen2023lossless,7,8,fang20223dac,17}.
Despite some achievements in geometry compression, in terms of attribute compression, point-based compression methods have difficulty leveraging excellent feature extraction operators, such as convolution, or voxel-based compression networks are solely composed of convolution stacks \cite{7,8}. These could potentially lead to the network struggling to remove redundancy among highly correlated voxels, resulting in suboptimal compression efficiency.}

	In this paper, we propose a Transformer and Sparse Convolution based Point Cloud Attribute Compression (TSC-PCAC) to improve the coding efficiency. Main contributions are	
	\begin{itemize}
\item We propose a deep learning based point cloud attribute compression framework, called TSC-PCAC, which combines the Transformer and Sparse Convolutional Module (TSCM) based autoencoder and a TSCM optimized channel context module to improve compression efficiency.		
\item We propose a two-stage TSCM that jointly models global and local features to reduce data redundancy. The first stage focuses on capturing the dependencies and feature representations of local regions of the point clouds, while the second stage captures global features by combining spatial and channel pooling operations.
		\item 
		We design a channel context module based upon the TSCM, which divides quantified latent representations into multiple groups. It improves prediction accuracies of probability distributions by leveraging decoded groups.
		
	\end{itemize}
	The paper is organized as follows. Section \ref{Related} presents the related work. {Section \ref{section2}} presents motivations and the proposed TSC-PCAC. Section \ref{section3} presents the experimental results and analysis. Finally, Section \ref{section4} draws the conclusions.

	\section{Related Work}
		\label{Related}
		
	\subsection{Traditional Point Cloud Compression}
	
	Traditional point cloud compression methods are mainly divided into two categories: octree based and projection based.
	 Octree can efficiently and losslessly represent point clouds, where adjacent nodes exhibit high spatial correlation for point cloud compression. G-PCC \cite{mammou2019g} employs an octree-based compression method.
	 The attribute compression of G-PCC integrates three key compression methods, which are Region-Adaptive Hierarchical Transform (RAHT) \cite{2}, predictive transform, and lifting transform \cite{1}. RAHT is a variant of the Haar wavelet transform, using lower octree levels to predict values at the next level. Predictive transform generates different levels of detail based on distance and performs encoding using the order of detailed levels. Lifting transform is performed on top of predictive transform with an additional updating operation \cite{graziosi2020overview,10050256}.
	  Pavez \textit{et al.} \cite{pavez2021multi} utilized a Regionally Adaptive Graph Fourier Transform (RAGFT) to obtain low-pass and high-pass coefficients, and then used the decoded low-pass coefficients to predict the high-pass coefficients.
	  Gao \textit{et al.} \cite{10184956} exploited the statistical correlation between color signals to construct the Laplace matrix, which overcame the interdependence between graph transform and entropy coding and improved the coding efficiency.
	 V-PCC is a projection-based compression method.
	It divides the point cloud sequences into 3D patches and projects them into 2D geometry and attribute videos, which are then smoothed and compressed by the conventional 2D codecs. Zhang \textit{et al.} \cite{zhang2023perceptually} proposed a perceptually weighted Rate-Distortion (RD) optimization scheme for V-PCC, where perceptual distortion of point clouds \cite{WuTCSVT2021} was considered in objectives of attribute and geometry video encoding. These are traditional coding schemes exploiting spatial-temporal correlations, symbol and visual redundancies.
	\subsection{Learning-based Image Compression}
Deep learning and neural networks have achieved great {success} in solving vision tasks \cite{10254206,9417228,xing2023gqe}. Learning-based image compression \cite{ball,13,12,zou2022devil,minnen2020channel,6} has been hotspot in recent years, which has surpassed the traditional image compression methods, such as JPEG2000 and VVC Intra coding. The framework of learned image compression mainly transforms an image into a compact representation, which then undergoes quantization and entropy encoding for transmission. At the decoder side, the output image is synthesized from the decoded compact representation from entropy decoded bit stream. The entire end-to-end compression framework is optimized using RD loss as the objective function \cite{ball}. Ball{\'e} \textit{et al.} \cite{ball} applied a hyper-prior module into the Variational AutoEncoder (VAE) compression framework. Minnen \textit{et al.} \cite{6} proposed a context module exploiting autoregressive and hierarchical priors to improve compression efficiency, which was further extended to model channel-wise context \cite{minnen2020channel}. Tang \textit{et al.} \cite{12} utilized graph attention mechanism and asymmetric convolution to capture long-distance dependencies. Zou \textit{et al.} \cite{zou2022devil} proposed to combine local perceptual attention mechanism with global-related feature learning. Then, a window-based image compression was proposed to exploit spatial redundancies. Liu \textit{et al.} \cite{13} proposed a mixed transformer-{Convolutional Neural Network (CNN)} module, which was incorporated to optimize channel context module for learned image compression.
However, unlike the regular distributed 2D image, a point cloud is dispersed in 3D space in an unstructured and sparse manner. These 2D image compression methods cannot be directly applied to encode 3D point clouds.
	\subsection{Learning-based Point Cloud Compression}
To compress point clouds effectively, a number of learning-based compression methods have been proposed, which are categorized into learned geometry and attribute compression.

Generally, end-to-end learned Point Cloud Geometry Compression (PCGC) can be categorized into point-based methods  \cite{huang20193d,5} and voxel-based methods \cite{4,15,16}. The point-based methods process point clouds in 3D perspective. Huang \textit{et al.} \cite{huang20193d} developed a hierarchical autoencoder that incorporated multi-scale losses to achieve detailed reconstruction. Gao \textit{et al.} \cite{5} utilized local graphs for extracting neighboring features and employed attention-based methods for down-sampling points. These point-based methods struggle to efficiently aggregate spatial information. Voxel-based methods require an initial voxelization of point clouds before input to the compression network. Song \textit{et al.} \cite{song2023efficient} designed a grouping context structure and a hierarchical attention module for the entropy model of octrees, which supported parallel decoding. Wang \textit{et al.} \cite{16} extended the Inception-Resnet block \cite{10} to 3D domain, and then incorporated it into the VAE based compression network to address the gradient vanishing problem from going deeper. However, the employed dense convolution {required} intensive computations and large memory, which were unaffordable in processing large-scale point clouds. Wang \textit{et al.} \cite{4} introduced sparse convolution to accelerate computations and reduce memory cost. Additionally, a hierarchical reconstruction method was proposed to enhance compression performance. Liu \textit{et al.} \cite{15} introduced local attention mechanisms in PCGC, which achieved promising coding efficiency and demonstrated the potential of attention mechanisms. {Qi \textit{et al.} \cite{qi2024variable} proposed a variable rate PCGC framework to realize multiple rates by tuning the features.}  Akhtar \textit{et al.} \cite{10380494} proposed to predict a latent potential representation of the current frame from those of previous frames by using a generalized sparse convolution. {Zhou  \textit{et al.} \cite{zhou2024dynamic} proposed using transformers to capture temporal context information and reduce temporal redundancy. They also used a multi-scale approach to capture the temporal correlations of multi-frame point clouds over both short and long ranges.}  These methods are proposed to encode the point cloud geometry. However, the point cloud attribute is attached to the geometry, which is significantly different from the geometry.

To compress the point cloud attribute effectively, a number of Point Cloud Attribute Compression (PCAC) {methods} were developed, which can be categorized as lossless \cite{wang2023lossless,nguyen2023lossless} and lossy compression \cite{7,8,fang20223dac,17,isik2022lvac}. In lossless compression, Nguyen \textit{et al.} \cite{nguyen2023lossless} constructed the context information between the color channels to reduce redundancy among RGB components. Wang \textit{et al.} \cite{wang2023lossless} generated multi-scale attribute tensor and encoded the current scale attribute conditioned on lower scale attribute. In addition, it further reduced spatial and channel redundancy by grouping spatial and color channels. {However, lossless compression can hardly achieve high compression ratio, which motivates lossy compression. }
In lossy compression, {Isik \textit{et al.} \cite{isik2022lvac}  modeled point cloud attribute as a volumetric function and then compressed the parameters of the volumetric function to achieve attribute compression.  However, fitting the function leads to high coding time complexity.} Fang \textit{et al.} \cite{fang20223dac} employed the RAHT for initial encoding and designed a learning-based approach to estimate entropy. {Pinheiro \textit{et al.} \cite{17} proposed a Normalized Flow Point Cloud Attribute Compression (NF-PCAC) method.} It incorporated strictly reversible modules to improve the reversibility of data reconstruction, which extended the upper bound of coding performance. {Similarly, Lin  \textit{et al.} \cite{lin2023sparse} utilized the normalized flow paradigm by proposing the Augmented Normalizing Flow (ANF) model, in which additional adjustment variables were introduced}. Zhang \textit{et al.} \cite{zhang2023scalable} used two-stage coding, the point cloud was roughly coded using G-PCC in the first stage and it was augmented with a network in the second stage.


There are two more advanced PCAC methods based on the VAE framework, which are the Deep PCAC (Deep-PCAC) \cite{7} and the Sparse convolution-based PCAC (Sparse-PCAC) \cite{8}. In Deep-PCAC, a PointNet++ \cite{9} based feature extraction was proposed to provide a larger receptive field for aggregating neighboring features. {However, since the Deep-PCAC processes irregular points directly, it cannot utilize convolution operators on regular voxels to extract local features\cite{10}.} {The latter Sparse-PCAC processed voxelized	point clouds, which used convolution to aggregate neighboring	features and sparse convolution to reduce the computation	and memory requirements \cite{11}}. However, Sparse-PCAC used stacked convolutional layers with fixed kernel weights, degrading the model's ability of focusing highly relevant voxels.


\section{The proposed TSC-PCAC}
\label{section2}
\subsection{Motivations and Analysis}
	Point cloud is sparsely distributed in 3D space, which suits for sparse processing in neighboring feature extraction. In addition, attention mechanisms that can capture correlations between voxels are also applicable to the point cloud compression. Therefore, the properties of sparse convolution and local attention are analyzed for the attribute coding.
	{
\subsubsection{Importance of Sparse Convolution}
	Dealing with large-scale dense point clouds requires a large number of parameters and computational cost of 3D convolutional networks, which is challenging in constructing effective compression networks for large-scale point clouds.} {Sparse convolution is effective in reducing the computational complexity by exploiting the sparse and vacant points in point clouds.} The general representation of 3D convolution is
\begin{equation}
	\boldsymbol{f}_{\boldsymbol{u}}^{\text {out }}=\sum_{\boldsymbol{i}} \mathbf{W}_{\boldsymbol{i}} \boldsymbol{f}_{\boldsymbol{u}+\boldsymbol{i}}^{\text{in}},
\end{equation}
where $\boldsymbol{f}_{\boldsymbol{u}}^{\text {out}}$ represents the output feature at 3D spatial coordinate $\boldsymbol{u}$ and $\boldsymbol{f}_{{\boldsymbol{u}}+{\boldsymbol{i}}}^{\mathrm{in\ }}$ represents the input feature at the position  $\boldsymbol{u+i}$ , where $\boldsymbol{i}$ is an offset of $\boldsymbol{u}$ in 3D space. $\mathbf{W}_{\boldsymbol{i}}$ represents the convolution kernel weight corresponding to offset $\boldsymbol{i}$.
				\begin{figure}[t]
		\centering
		\includegraphics[width=1\columnwidth]{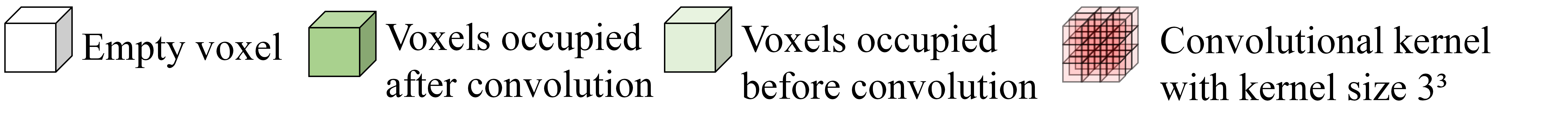}
		\subfigure[] {\label{fig9a}
			\includegraphics[width=0.45\columnwidth]{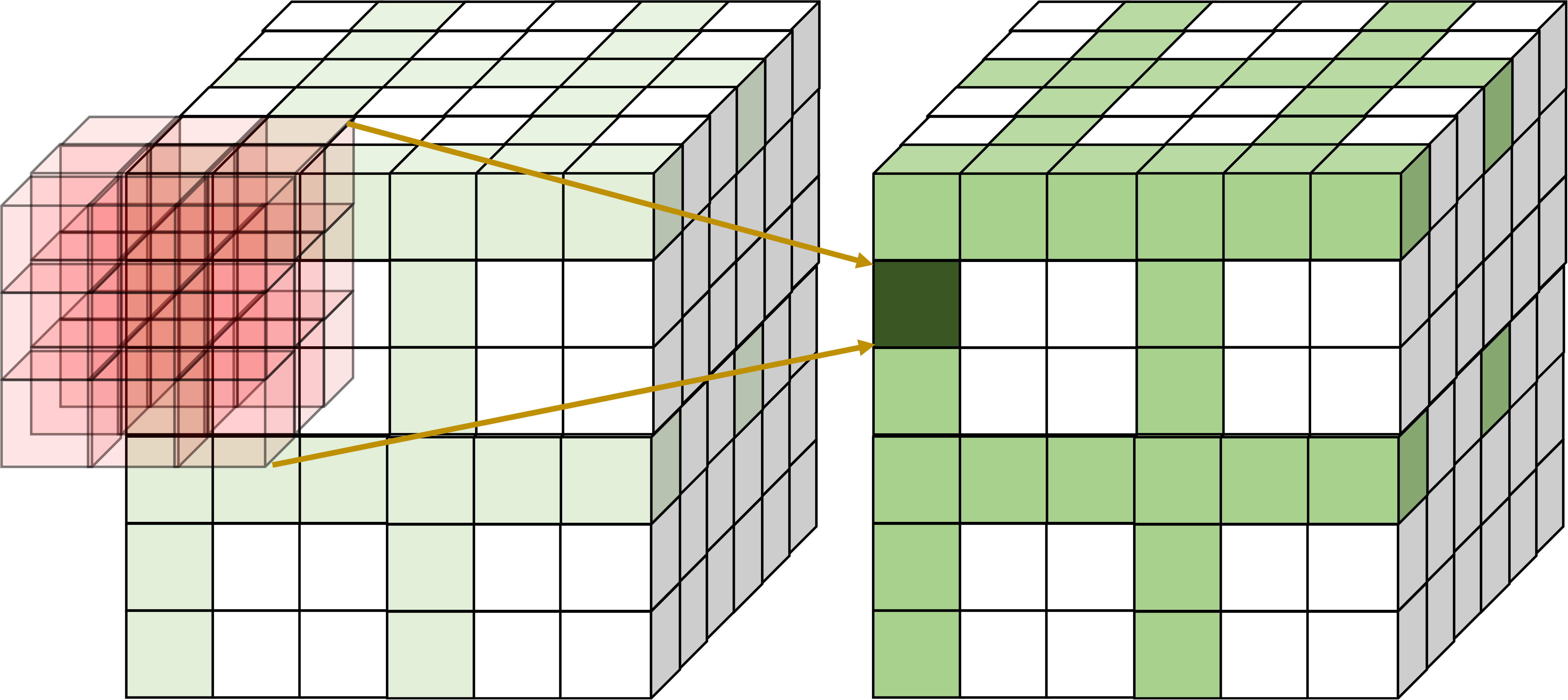}}
		\subfigure[] {\label{fig9b}
			\includegraphics[width=0.45\columnwidth]{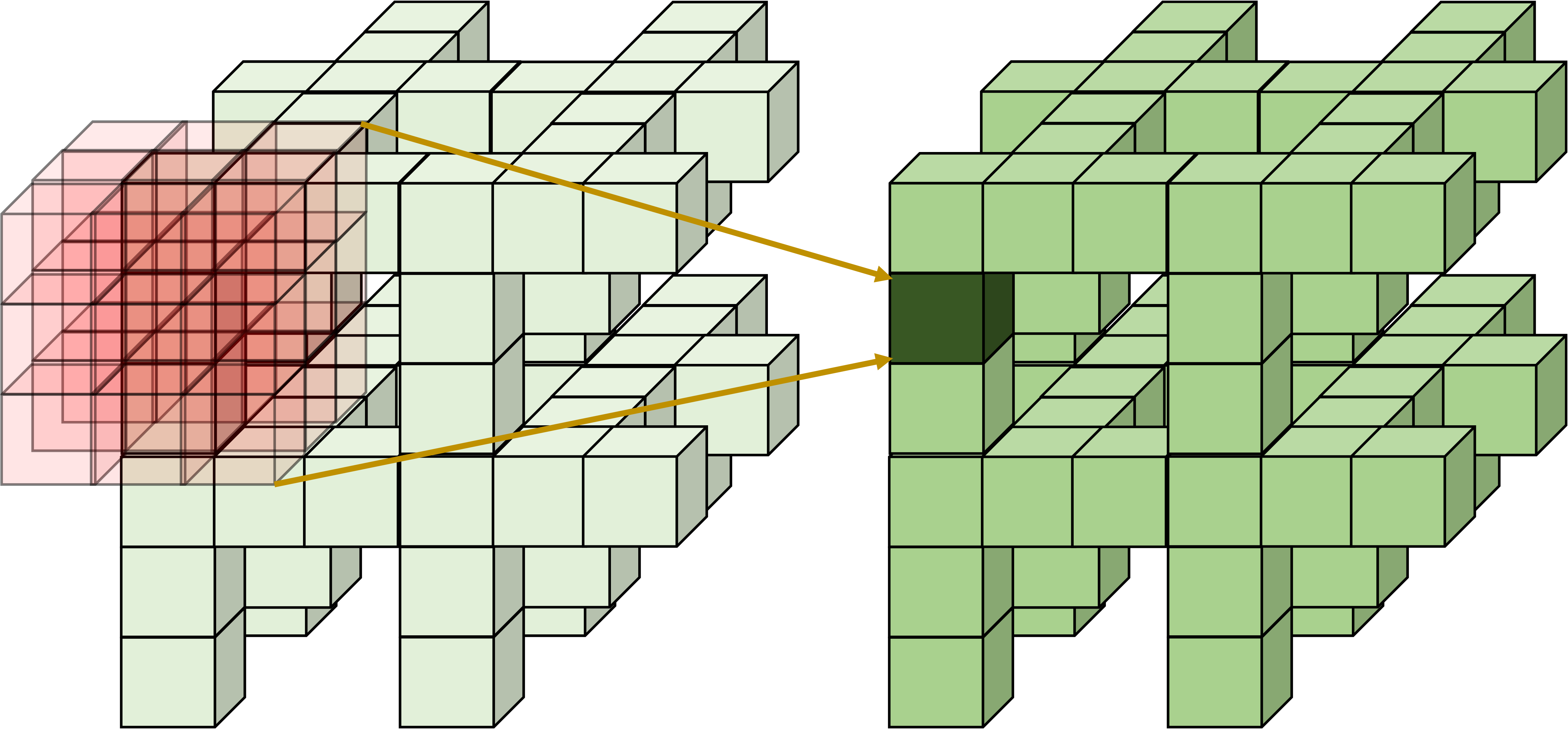}}
		\caption{Differences between dense and sparse convolution with kernel size $3\times3\times3$. (a) dense convolution, (b) sparse convolution.}
		\label{fig9}
	\end{figure}
	\begin{figure}[t]
		{	\centering
			\subfigure[]{\includegraphics[scale=0.45]{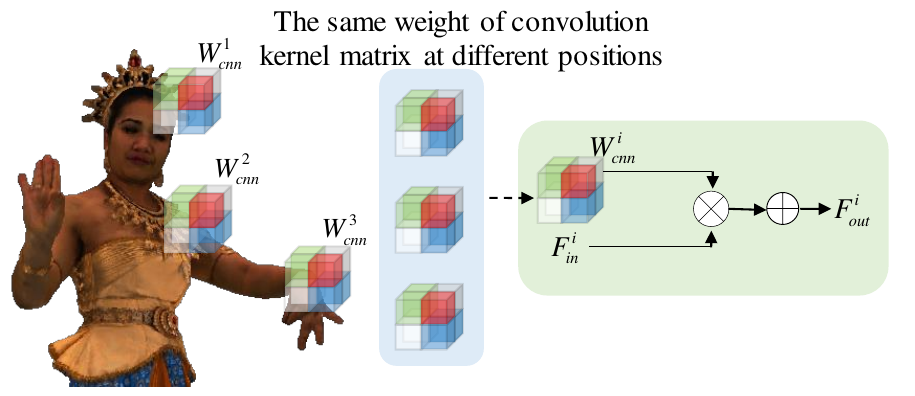}%
				\label{fig10a}}
			\subfigure[]{\includegraphics[scale=0.45]{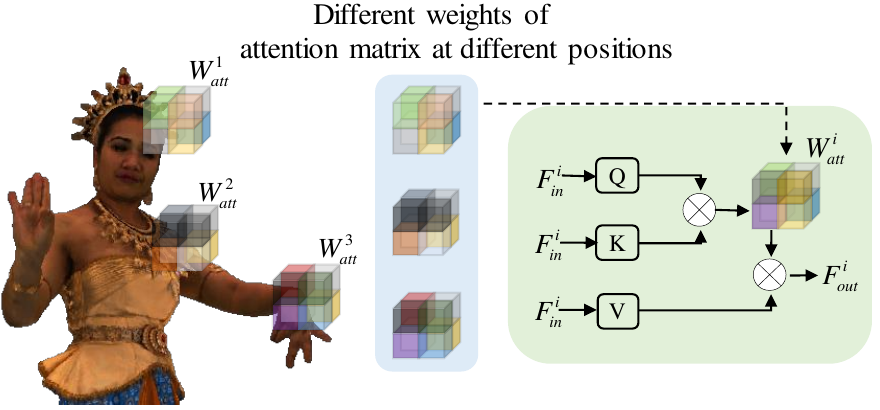}%
				\label{fig10b}}
			\caption{Weight differences between convolution and local attention for voxels. (a) convolution in CNN. (b) local attention in Transformer, {where $\mathbf{Q},\mathbf{K},\mathbf{V}$ denote three linear layers of query, key and value, respectively.} }
			\label{fig10}
		}
	\end{figure}
	For spatial scale-invariant dense convolution, $\boldsymbol{i}$ is an offset selected from the list of offsets $\mathcal{V}^{3}(K)$ in a 3D hypercube with center $\boldsymbol{u}$, where $K$ is the size of one dimension of the convolution kernel.
	More intuitively, Fig. \ref{fig9} illustrates the difference between sparse convolution and dense convolution. For a convolution operation with kernel size $3^3$, dense convolution stores both occupied and vacant voxels. During the convolution, it acts on all 27 voxels covered by the convolution kernel.
	However, for spatial scale-invariant sparse convolution, $\boldsymbol{i}$ is an offset selected from the set of offsets $\mathcal{N}^{{3}}(\boldsymbol{u}, \boldsymbol{C}^{\text {in}})$, which is defined as the set of offsets from the current center $\boldsymbol{u}$ that exists in the set $\boldsymbol{C}^{\text {in }}$. $\boldsymbol{C}^{\text{in}}$ are predefined input coordinates of sparse tensors that only include occupied voxels. As depicted in Fig. \ref{fig9b}, sparse convolution only stores and acts on the occupied voxels  \cite{11}. In addition, sparse convolution does not perform convolution with vacant voxels at the center. Compared with the dense convolution, sparse convolution {significantly} reduces the memory cost and computational complexity, which enables the design of more powerful compression networks for large-scale point clouds.

	\begin{figure*}[!t]
		\centering

		\includegraphics[width=6in]{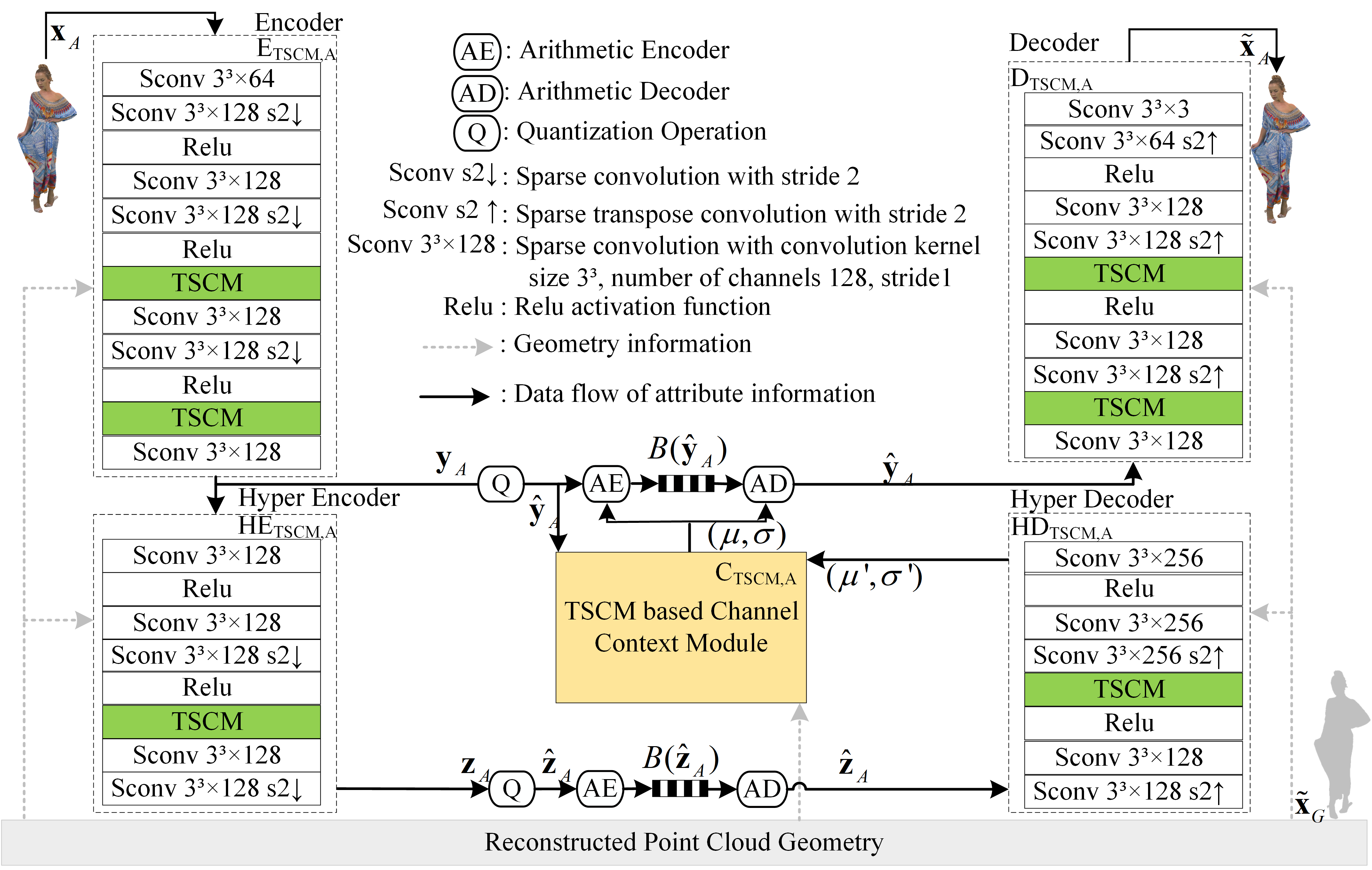}
		\caption{Framework of the proposed TSC-PCAC, where green and yellow rectangles are proposed TSCM and TSCM based channel context model.}
		\label{fig1}
	\end{figure*}	
	
	\subsubsection{Importance of Local Attention}
	Fig. \ref{fig10a} shows the learned weights of the convolution kernels in point cloud processing. {There is only one set of weights in each convolution of CNN and it traverses the entire point cloud using a sliding convolution kernel to extract features, which leads to learn a universal convolution kernel for different point cloud regions.} Obviously, the properties of the regions/voxels vary from regions. Therefore, it is not the optimal solution to treat all voxels equally regardless {of} the region contents. In contrast, the self-attention in transformer uses the input to compute the attention matrix and model the dependencies of different voxels \cite{25}. As shown in Fig. \ref{fig10b}, local attention operator in transformers computes the attention weights of the input voxels with respect to other voxels, which {is} able to capture correlations between voxels. In this case, the correlations between voxels are captured using the self-attention, which is to model the redundancy between voxels and reduce the coding bitrate.
	Moreover, comparing with the convolution, transformer utilizes self-attention mechanisms to establish mutual dependencies between voxels, which enhances the capability of feature representation. However, transformer lacks an effective inductive bias, compromising the generalization performance, especially in 3D point cloud processing. Combining transformer with sparse convolution can improve the capabilities of feature representation and generalization.
	
	\subsection{Framework of the TSC-PCAC}
	Given a point cloud $\mathbf{x}_{pc}$, a learned point cloud compression framework to encode and decode $\mathbf{x}_{pc}$ can be expressed as
		{
	\begin{equation}
		\label{con:inventoryflow}
		\begin{cases}
			\begin{aligned}
				&B(\hat{\mathbf{y}}_{pc}) =AE(Q( E_{pc}(	\mathbf{x}_{pc}))) \\
				&\tilde{	\mathbf{x}}_{pc} = D_{pc}(AD(B(\hat{\mathbf{y}}_{pc})))
			\end{aligned}
		\end{cases},
	\end{equation}}where $E_{pc}(\cdot)$ and $D_{pc}(\cdot)$ represent the point cloud encoder and decoder, respectively. $AE(\cdot)$ and $AD(\cdot)$ respectively represent arithmetic encoder and decoder. $\mathbf{x}_{pc}$ and $\tilde{\mathbf{x}}_{pc}$ represent the source and the reconstructed point clouds, respectively. $\hat{\mathbf{y}}_{pc}$ and $B(\hat{\mathbf{y}}_{pc})$ represent the compact latent representations after quantization and the bitstream containing the $\hat{\mathbf{y}}_{pc}$ information. $Q(\cdot)$ is a quantization operation.
	
	A point cloud $\mathbf{x}_{pc}$ consists of geometry $\mathbf{x}_{G}$ and attribute $\mathbf{x}_{A}$, i.e., $\mathbf{x}_{pc}=[\mathbf{x}_{A},\mathbf{x}_{G}]$. Since attribute attaches to geometry and is compressed after geometry \cite{mammou2017video, mammou2019g}. Eq.(\ref{con:inventoryflow}) is rewritten as 	{
	\begin{equation}
		\begin{cases}
			\begin{aligned}
				&B(\hat{\mathbf{y}}_{G}) =AE(Q( E_{G}(	\mathbf{x}_{G}))) \\
				&\tilde{	\mathbf{x}}_{G} = D_{G}(AD(B(\hat{\mathbf{y}}_{G})))\\
				&B(\hat{\mathbf{y}}_{A}) =AE(Q( E_{A}(	\mathbf{x}_{A},\tilde{	\mathbf{x}}_{G}))) \\
				&\tilde{	\mathbf{x}}_{A} = D_{A}(AD(B(\hat{\mathbf{y}}_{A})),\tilde{	\mathbf{x}}_{G})
			\end{aligned}
		\end{cases},
	\end{equation} }where $\tilde{\mathbf{x}}_{G}$ and $\tilde{\mathbf{x}}_{A}$ represent the reconstructed point cloud geometry and attribute.  $\hat{\mathbf{y}}_{G}$ and $\hat{\mathbf{y}}_{A}$ represent the quantified latent representations of geometry and attribute. $E_{G}(\cdot)$ and $D_{G}(\cdot)$ represent point cloud geometry encoder and decoder. $E_{A}(\cdot)$ and $D_{A}(\cdot)$ represent the point cloud attribute encoder and decoder. $B(\hat{\mathbf{y}}_{G})$ and $B(\hat{\mathbf{y}}_{A})$ represent the bitstream containing the $\hat{\mathbf{y}}_{G}$ and $\hat{\mathbf{y}}_{A}$ information.  Attribute $\mathbf{x}_{A}$ is on the basis of geometry ${\mathbf{x}}_{G}$, thus geometry is compressed before the attribute, and then the attribute is compressed with the reconstructed geometry information $\tilde{\mathbf{x}}_{G}$ as an input.
	
	In this work, suppose that point cloud geometry has already been compressed, we focus on optimizing the PCAC. Consequently, our goal is to optimize $E_A(\cdot)$ and $D_A(\cdot)$, as well as the entropy model required in arithmetic codec while encoding and decoding $\hat{\mathbf{y}}_A$. We propose an efficient TSC-PCAC framework based on the VAE, as shown in Fig. \ref{fig1}, in which TSCM and a TSCM based context module are proposed to improve the point cloud coding. The TSC-PCAC combines the inductive bias capabilities and memory-saving advantages of sparse convolution, and also leverages the ability of transformers in capturing correlations between voxels.
	The TSC-PCAC consists of a pair of learning-based encoder/decoder and a pair of entropy model, which is	
	\begin{equation}
		\begin{cases}
			\begin{aligned}
				&B(\hat{\mathbf{y}}_{A})=AE(Q(E_{{TSCM},A}(\mathbf{x}_A,\tilde{\mathbf{x}}_{G} ; \boldsymbol{\phi}))) \\
				&\tilde{\mathbf{x}}_A=D_{{TSCM},A}(AD(B(\hat{\mathbf{y}}_{A})),\tilde{\mathbf{x}}_G ; \boldsymbol{\theta})\\
				&\{\boldsymbol{\mu}, \boldsymbol{\sigma}\}=Entropy_A(\mathbf{y}_A,\tilde{\mathbf{x}}_{G})
			\end{aligned}
		\end{cases},
	\end{equation}
where ${\mathbf{y}}_{A}$ and $\hat{\mathbf{y}}_A$ are the original and quantized latent representations of attribute, respectively. $E_{{TSCM},A}(\cdot)$ and $D_{{TSCM},A}(\cdot)$ are TSCM-optimized attribute encoder and decoder with parameters $\boldsymbol{\phi}$ and $\boldsymbol{\theta}$, respectively. $Entropy_A(\cdot)$ represents the entropy model for $\hat{\mathbf{y}}_A$. The mean $\boldsymbol{\mu}$ and scale $\boldsymbol{\sigma}$ from the entropy model are used for the arithmetic codec $AE(\cdot)$ and $AD(\cdot)$. Specifically, the entropy model consists of a hyper-prior encoder-decoder and context module. In estimating the probability distribution of $\hat{\mathbf{y}}_A$, the initial mean and scale are firstly achieved by a hyper-prior encoder-decoder, and then refined by a context module.
	The TSCM based channel context model and entropy coding can be represented as 	{
	\begin{equation}
		\begin{cases}
			\begin{aligned}	
				&B(\hat{\mathbf{z}}_A)=AE(Q(HE_{{TSCM},A}\left(\mathbf{y}_A,\tilde{\mathbf{x}}_{G} ; \boldsymbol{\phi}_h\right))) \\
				&\{\boldsymbol{\mu^\prime}, \boldsymbol{\sigma^\prime}\}=HD_{{TSCM},A}\left(AD(B(\hat{\mathbf{z}}_A)),\tilde{\mathbf{x}}_G ; \boldsymbol{\theta}_h\right) \\
				&\{\boldsymbol{\mu}, \boldsymbol{\sigma}\}=C_{{TSCM,A}}\left(\boldsymbol{\mu^\prime}, \boldsymbol{\sigma^\prime},\hat{\mathbf{y}}_A,\tilde{\mathbf{x}}_G ; \boldsymbol{\theta}_c\right)
			\end{aligned}
		\end{cases},
	\end{equation} }where $B(\hat{\mathbf{z}}_A)$ represent the bitstream containing the quantized hyperprior representations $\hat{\mathbf{z}}_A$ information. $HE_{{TSCM},A}(\cdot)$ and $HD_{{TSCM},A}(\cdot)$ represent the hyperprior encoder and decoder with parameters $\boldsymbol{\phi}_h$ and $\boldsymbol{\theta}_h$, respectively. $\boldsymbol{\mu^\prime}$ and $\boldsymbol{\sigma^\prime}$ represent initial mean and scale of $\hat{\mathbf{y}}$. $C_{{TSCM},A}(\cdot)$ denotes the TSCM based channel context module with parameters $\boldsymbol{\theta}_c$, which establishes relationships between channels to provide contextual information for higher compression efficiency.

	To enhance the network's capability of allocating more importance to voxels with higher relevance and obtaining a compact representation, TSCM blocks are placed in the encoder $E_{{TSCM},A}(\cdot)$ and decoder $D_{{TSCM},A}(\cdot)$. Specifically, due to memory constraints, only two TSCM blocks are used in the encoder and decoder. Similarly, one TSCM block is employed in the hyper-prior encoder and decoder. Since the attribute is attached to geometry, the geometry information is utilized in the entire process of attribute coding, shown as gray arrows in Fig. \ref{fig1}. For simplicity, geometry variables are not explicitly indicated in subsequent diagrams.
	
	\begin{figure*}[!t]
		\centering
		\subfigure[] {
			\label{fig2a}
			\includegraphics[scale=0.75]{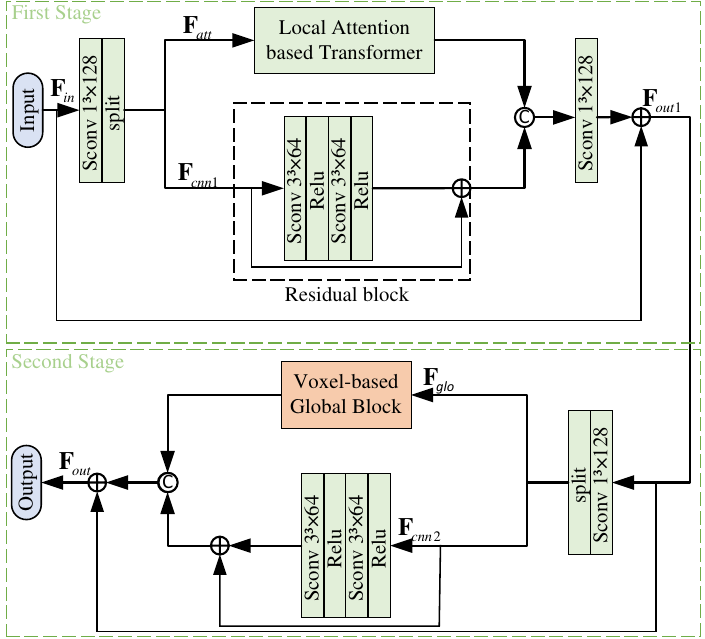}
		}
		\subfigure[] {
			\label{fig2b}
			\includegraphics[scale=0.75]{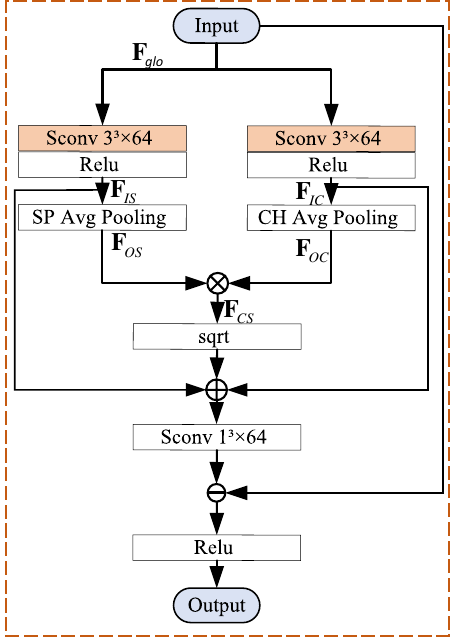}
		}
		\caption{The structure of the TSCM and its key module. (a) TSCM, (b) Voxel-based Global Block. {The `split' operation performs channel-wise splitting. `SP Avg Pooling' refers to spatial average pooling, which applies global pooling operation across spatial dimensions, and `CH Avg Pooling' refers to channel-wise average pooling, which applies global pooling operation across channel dimensions. {The `sqrt' represents the square root operation on the input.}}}
		\label{fig2}
	\end{figure*}

	\subsection{Architecture of TSCM}
	Transformer models the correlations dependencies between voxels through the self-attention mechanism, which has the potential to enable the network to focus on voxels of higher relevance. However, it may reduce the generalization ability.
		{Convolution, on the other hand, has a strong inductive bias and can achieve excellent generalization ability. Combining convolution and transformer can gain their advantages. However, when using the transformer to process large-scale color point clouds that {contain} millions of points, global attention is able to capture long-range dependencies at {the} cost of the unaffordable memory, while local attention {reduces} the memory cost at {the} cost of reducing the receptive field.} To take the advantages of these methods, we propose a two-stage TSCM, which consists of local attention based transformer \cite{25}, residual block, and voxel-based global block to increase the receptive field.
	
Fig. \ref{fig2a} shows the structure of the proposed TSCM. In the first stage, residual block and local attention based transformer are utilized to extract local features. {To take advantage of these features, we use a dual-branch structure to extract {features} separately and fused by concatenation.}
	Specifically,  an input $\mathbf{F}_{{in}}\in\mathbb{R}^{N\times128}$ performs sparse convolution with a kernel size of 1 and splits along the channel dimension, resulting in $\mathbf{F}_{{att}}\in\mathbb{R}^{N\times64}$ and $\mathbf{F}_{{cnn}1}\in\mathbb{R}^{N\times64}$. They are used as inputs for the local attention-based transformer and residual block, respectively. The output is then concatenated to restore the feature channel to the original dimension, i.e., 128. This fusion combines both two types of features to form local features  $\mathbf{F}_{{out}1}\in\mathbb{R}^{N\times128}$, which captures interdependencies among voxels and has effective inductive bias. These features are the output of the first stage and the input of the second stage.

	In the second stage, to alleviate the computation/memory cost and achieve the global features, the residual block and the voxel-based global block are employed. The global block in \cite{28} uses linear layers and global pooling to extract and sharpen global features. To further capture spatial neighboring redundancy, voxel-based global block is proposed by replacing the linear layer in global block with convolutional layers. Specifically, $\mathbf{F}_{ {glo}}\in\mathbb{R}^{N\times64}$ and $\mathbf{F}_{{cnn}2}\in\mathbb{R}^{N\times64}$ are input to the voxel-based global block and residual block, respectively. The voxel-based global block is shown in Fig. \ref{fig2b}. It utilizes global pooling to extract global spatial features $\mathbf{F}_{{OS}}\in\mathbb{R}^{1\times64}$ and global channel features $\mathbf{F}_{{OC}}\in\mathbb{R}^{N\times1}$. $\mathbf{F}_{{OC}}$ and $\mathbf{F}_{{OS}}$ undergo matrix multiplication to obtain global spatial-channel features $\mathbf{F}_{{CS}}\in\mathbb{R}^{N\times64}$. Subsequently, the global spatial-channel features are subtracted from the input features of voxel-based global block $\mathbf{F}_{ {glo }}$ to obtain sharpened global spatial-channel features. The outputs of the voxel-based global block and residual block are concatenated to restore the feature channel to the original dimension. This fusion combines both two types of features to form global features with a large receptive field and inductive bias capabilities.
	
	 In TSCM, the integration of these two stages effectively captures both global and local inter-point relevance for reducing data redundancy. {We use both} the residual block and the transformer with local self-attention to extract local features and correlations. Meanwhile, the global features are obtained through channel and spatial pooling within voxel-based global blocks. Finally, the features of TSCM are output as $\mathbf{F}_{\text{out}} \in \mathbb{R}^{N \times 128}$.

	\subsection{TSCM based Channel Context Module}
	In the Sparse-PCAC, the context module is a per-voxel context network, which processes voxels sequentially. While dealing with large-scale point clouds, this sequential processing leads to extremely high encoding and decoding delays. We propose a TSCM based channel context module to improve the network prediction probability distribution by constructing contextual information over quantized latent representations. {Features are processed sequentially along the channel dimension instead of voxels, which {reduces} the sequential number.}
	
\begin{figure*}[!t]
		\centering
		\subfigure[] {
			\label{fig3a}
			\includegraphics[scale=0.8]{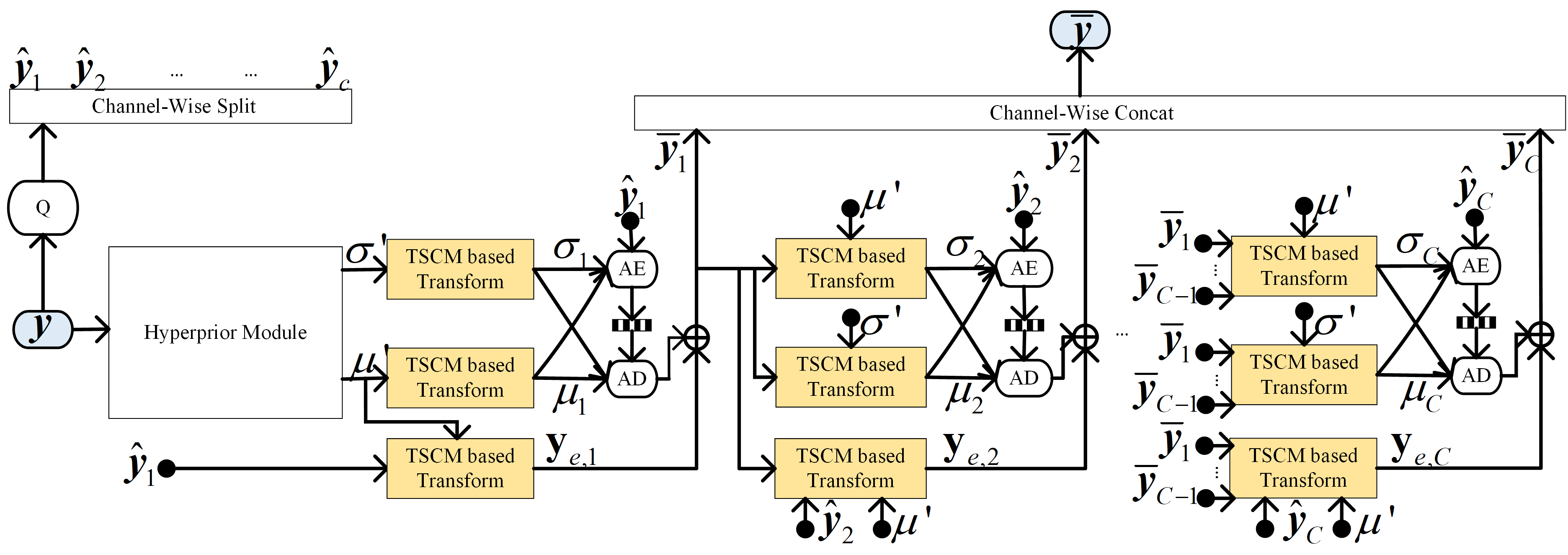}
		}
		\subfigure[] {
			\label{fig3b}
			\includegraphics[scale=1.1]{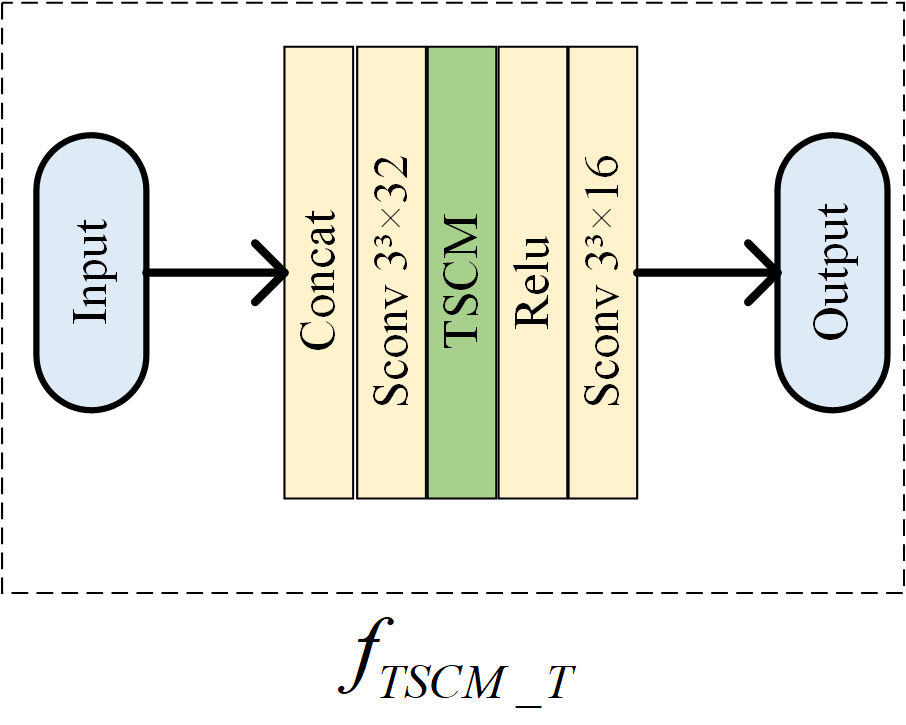}
		}
		\caption{Structure of the proposed TSCM based channel context module. (a) channel context module, (b) TSCM based transform.}
		\label{fig3}
	\end{figure*}

	The architecture of the proposed TSCM based channel context module is illustrated in Fig. \ref{fig3a}, where the hyper-prior module is detailed in Fig. \ref{fig1}. $\hat{\mathbf{y}}_i$ represents the latent representation of the $i$-th quantized slice, while $\overline{y}_i$ represents the latent representation of the $i$-th slice with the addition of quantization error estimation, where $i\in{\{1,2,3,\ldots,C\}}$. $C$ represents the number of equal divisions in the channel. Besides the mean and scale, we also employ a module to predict quantization error. Specifically, the channel context module processes each slice sequentially by dividing the quantized latent representation $\hat{\mathbf{y}}$ proportionally into slices along the channel. Predicting the mean of each slice depends on the previously decoded slices and the initial mean value. Similar process is applied to predict the scale. Since the obtained probability distribution can be used to decode the current slice, the quantization error can be predicted from the decoded current slice. It is worth noting that the network architectures for predicting the mean, scale, and quantization error are the same. So, we use the term `TSCM based Transform' for these three cases.
The detailed structure of the TSCM based transform is depicted in Fig. \ref{fig3b}, where inputs are concatenated and processed through sparse convolutions and TSCM block.	
Due to the computational complexity of TSCM, we reduce the number of feature channels to 32. The mathematical expressions of obtaining the scale $\boldsymbol{\sigma}_i$, mean $\boldsymbol{\mu}_i$, and quantization error ${\mathbf{y}}_{e,i}$ are
	\begin{equation}
		\begin{cases}
			\begin{aligned}
				\boldsymbol{\sigma}_i&=f_{\text {TSCM\_T }}\left(\overline{\mathbf{y}}_{<i}, \boldsymbol{\sigma}^{\prime};\boldsymbol{\theta}_{\sigma,i}\right) \\
				\boldsymbol{\mu}_i&=f_{\text {TSCM\_T }}\left(\overline{\mathbf{y}}_{<i}, \boldsymbol{\mu}^{\prime};\boldsymbol{\theta}_{\mu,i}\right) \\
				{\mathbf{y}}_{e,i}&=f_{\text {TSCM\_T }}\left(\overline{\mathbf{y}}_{<i},\hat{\mathbf{y}}_i ,\boldsymbol{\mu}^{\prime};\boldsymbol{\theta}_{y,i}\right)
			\end{aligned}
		\end{cases},
\label{Eq:F_TSCM}
	\end{equation}
	where $\boldsymbol{\mu}_i$ and $\boldsymbol{\sigma}_i$ represent the refined mean and scale of $\hat{\mathbf{y}}_i$. $ {\mathbf{y}}_{e,i}$ is the estimated quantization error. $\overline{\mathbf{y}}_{i}$ denotes the refined latent representation of group $i$, i.e., $\overline{\mathbf{y}}_{i}={\mathbf{y}}_{e,i}+\hat{\mathbf{y}}_i$. $\overline{\mathbf{y}}_{<i}$ indicates a set of $\overline{\mathbf{y}}_{j}$ where $j$ is smaller than $i$, $\overline{\mathbf{y}}_{<i} = \left\{\overline{\mathbf{y}}_1, \overline{\mathbf{y}}_2, \ldots, \overline{\mathbf{y}}_{i-1}\right\}$. {$f_{\text {TSCM\_T }}(\cdot)$ is the `TSCM based Transform' learning parameters of slices, which are $\boldsymbol{\theta}_{\sigma,i}$, $\boldsymbol{\theta}_{\mu,i}$, and $\boldsymbol{\theta}_{y,i}$, respectively.} The mean $\boldsymbol{\mu}_i$ and scale $\boldsymbol{\sigma}_i$ of $i$-th group are utilized for entropy encoding and decoding of the current group's $\hat{\mathbf{y}}_i$. The quantization error $\mathbf{y}_{e,i}$ predicted for each group is then employed to refine $\hat{\mathbf{y}}_i$. Based on Eq. \ref{Eq:F_TSCM} and Fig.\ref{fig3}, {the mean $\boldsymbol{\mu}_i$, scale $\boldsymbol{\sigma}_i$, and quantization error} ${\mathbf{y}}_{e,i}$ can be predicted by the TSCM based transform, which {improve} the channel context module in entropy coding.
	
	\subsection{Loss Function}
	
	The overall objective of point cloud compression is to minimize the distortion of the reconstructed point cloud at a given bitrate constraint. As a point cloud consists of geometry and attribute, the objective is to minimize
	{
	\begin{equation}
		\begin{gathered}
			L=R_A(\hat{\mathbf{y}}_A)+R_G(\hat{\mathbf{y}}_G)+R_A(\hat{\mathbf{z}}_A)+R_G(\hat{\mathbf{z}}_G)\\+	{\lambda_A} d_A(\mathbf{x}_A,\tilde{\mathbf{x}}_A)+	{\lambda_G} d_G(\mathbf{x}_G,\tilde{\mathbf{x}}_G),
		\end{gathered}
	\end{equation}
}where $d_A(\cdot)$ and $d_G(\cdot)$ are distortion measurements. $R_A(\hat{\mathbf{y}}_A)$ and $R_G(\hat{\mathbf{y}}_G)$ represent the bitrate required for transmitting the quantized attribute and geometry latent representations, $\hat{\mathbf{y}}_A$ and $\hat{\mathbf{y}}_G$. {Due to the hyper-codec, the total bit rate includes the rate of hyper-prior feature, i.e.,  $R_A(\hat{\mathbf{z}}_A)+R_G(\hat{\mathbf{z}}_G)$.} The trade-off between distortion and bitrate is controlled by $\lambda_A$ and $\lambda_G$. Furthermore, in this paper, we focus on the PCAC and assume the geometry information is shared and fixed at the encoder and decoder sides, Therefore, the objective for attribute coding is simplified to minimize
	\begin{equation}
		L_A=R_A(\hat{\mathbf{y}}_A)+R_A(\hat{\mathbf{z}}_A)+\lambda_A d_A(\mathbf{x}_A,\tilde{\mathbf{x}}_A).
	\end{equation}
In this paper, $d_A(\cdot)$ uses Mean Squared Error (MSE) loss in the YUV color space.
	\section{Experimental Results and Analyses}
	\label{section3}
	\subsection{Experimental Settings}

\begin{figure*}[!t]
	\centering
	\subfigure[] {
		\label{fig5a}
		\includegraphics[width=0.22\linewidth]{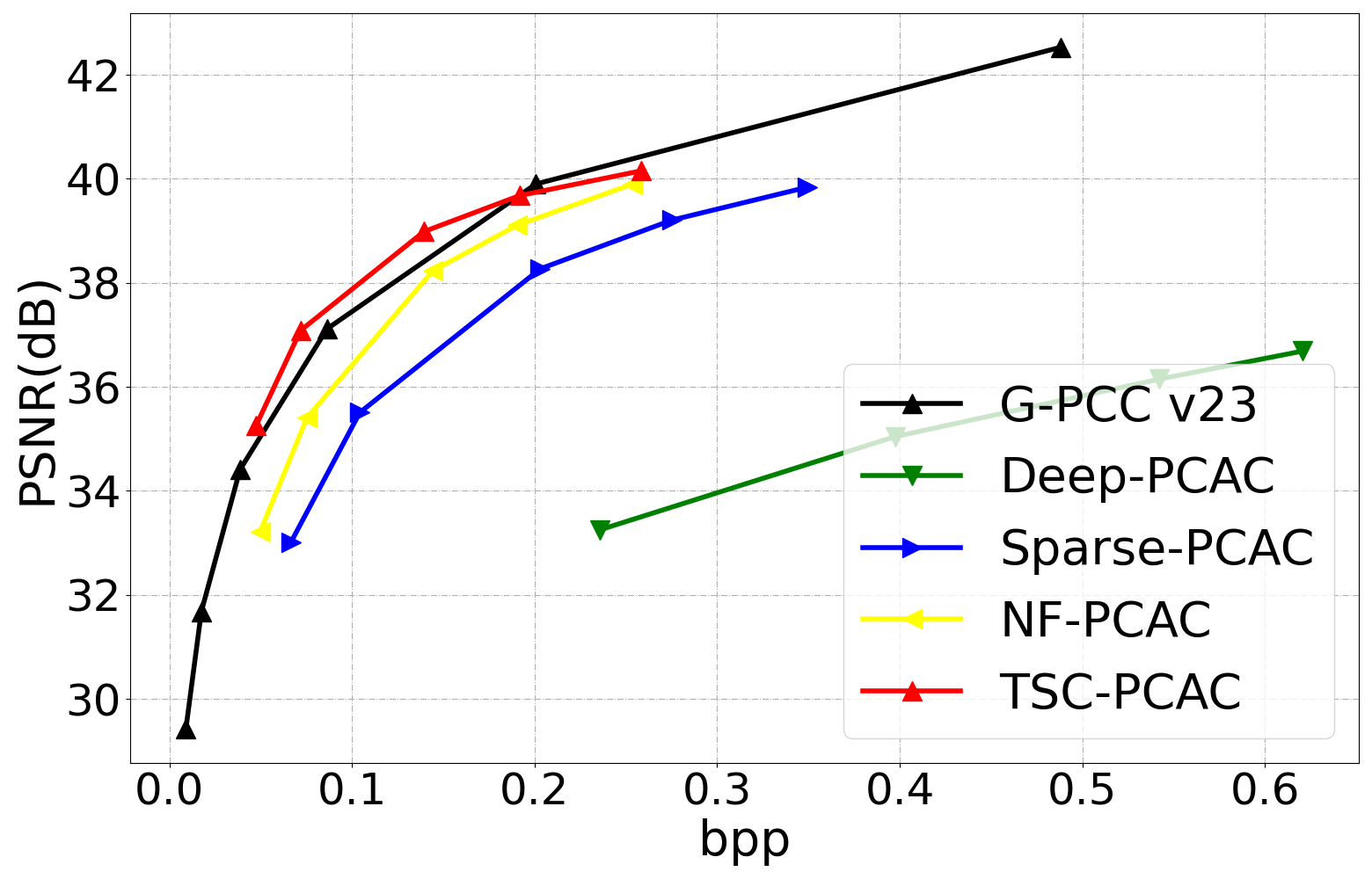}
	}
	\subfigure[] {
		\label{fig5b}
		\includegraphics[width=0.22\linewidth]{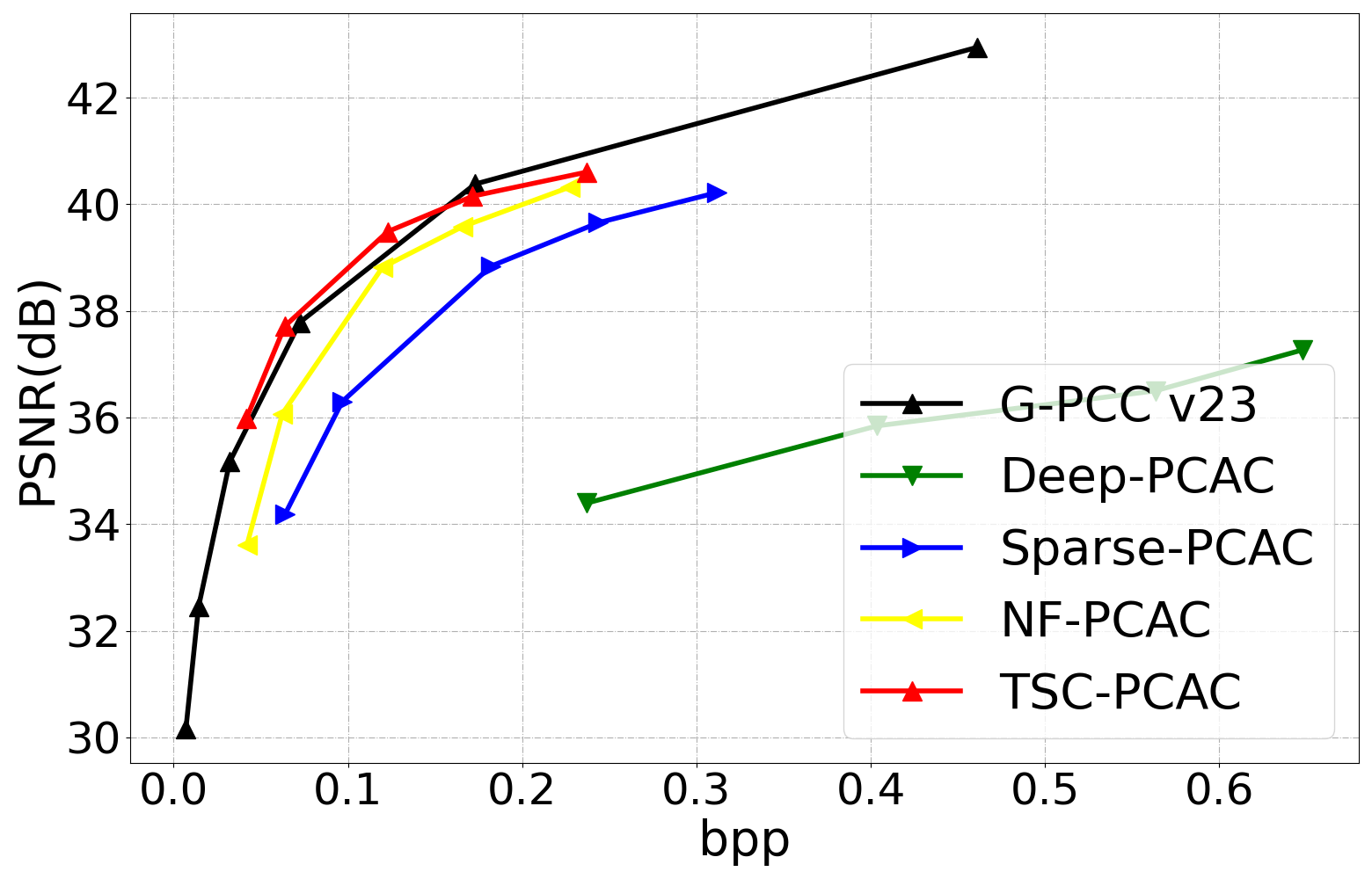}
	}
	\subfigure[] {
		\label{fig5c}
		\includegraphics[width=0.22\linewidth]{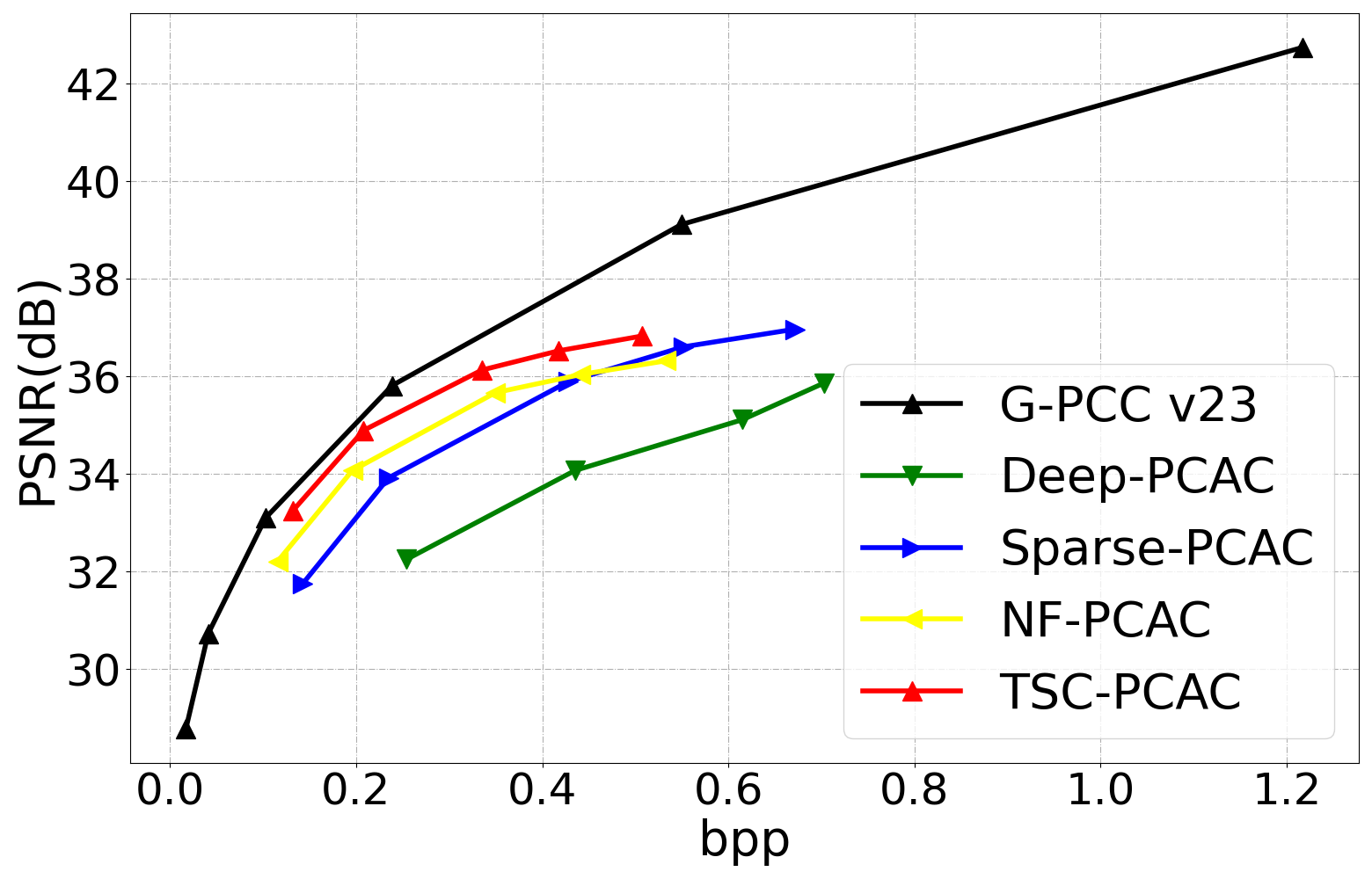}
	}
	\subfigure[] {
		\label{fig5d}
		\includegraphics[width=0.22\linewidth]{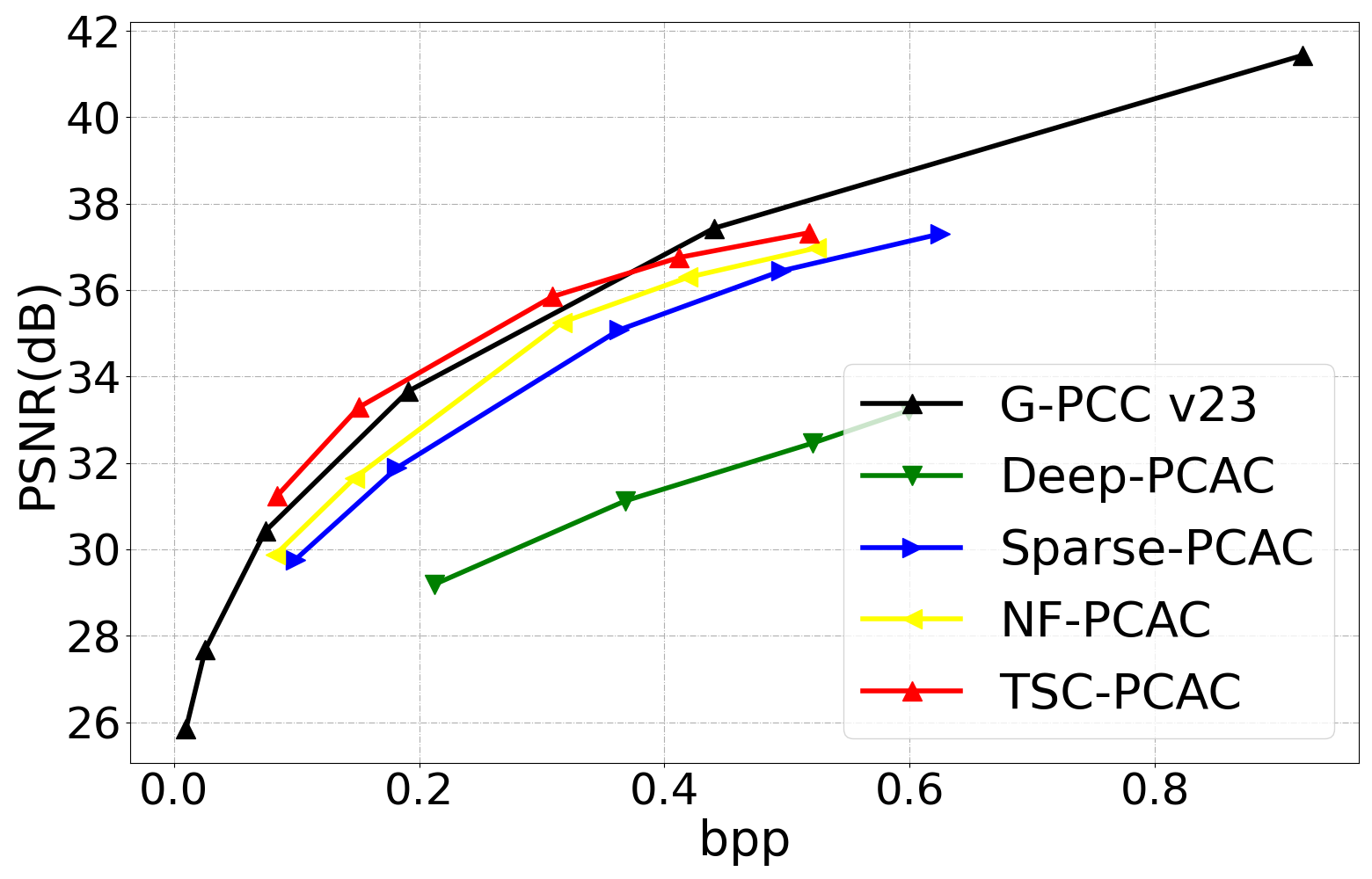}
	}
	
	\subfigure[] {
		\label{fig5e}
		\includegraphics[width=0.22\linewidth]{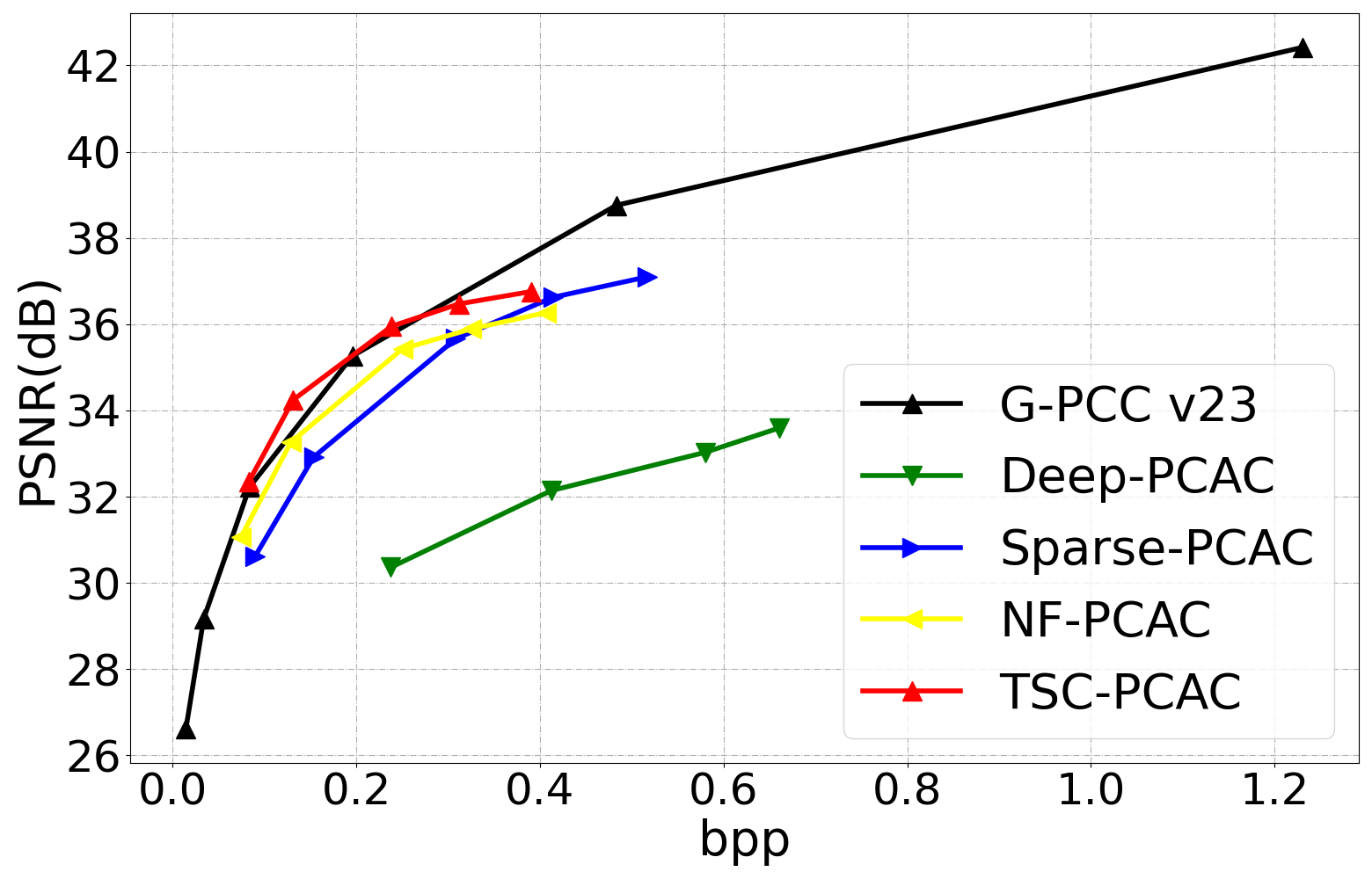}
	}
	\subfigure[] {
		\label{fig5g}
		\includegraphics[width=0.22\linewidth]{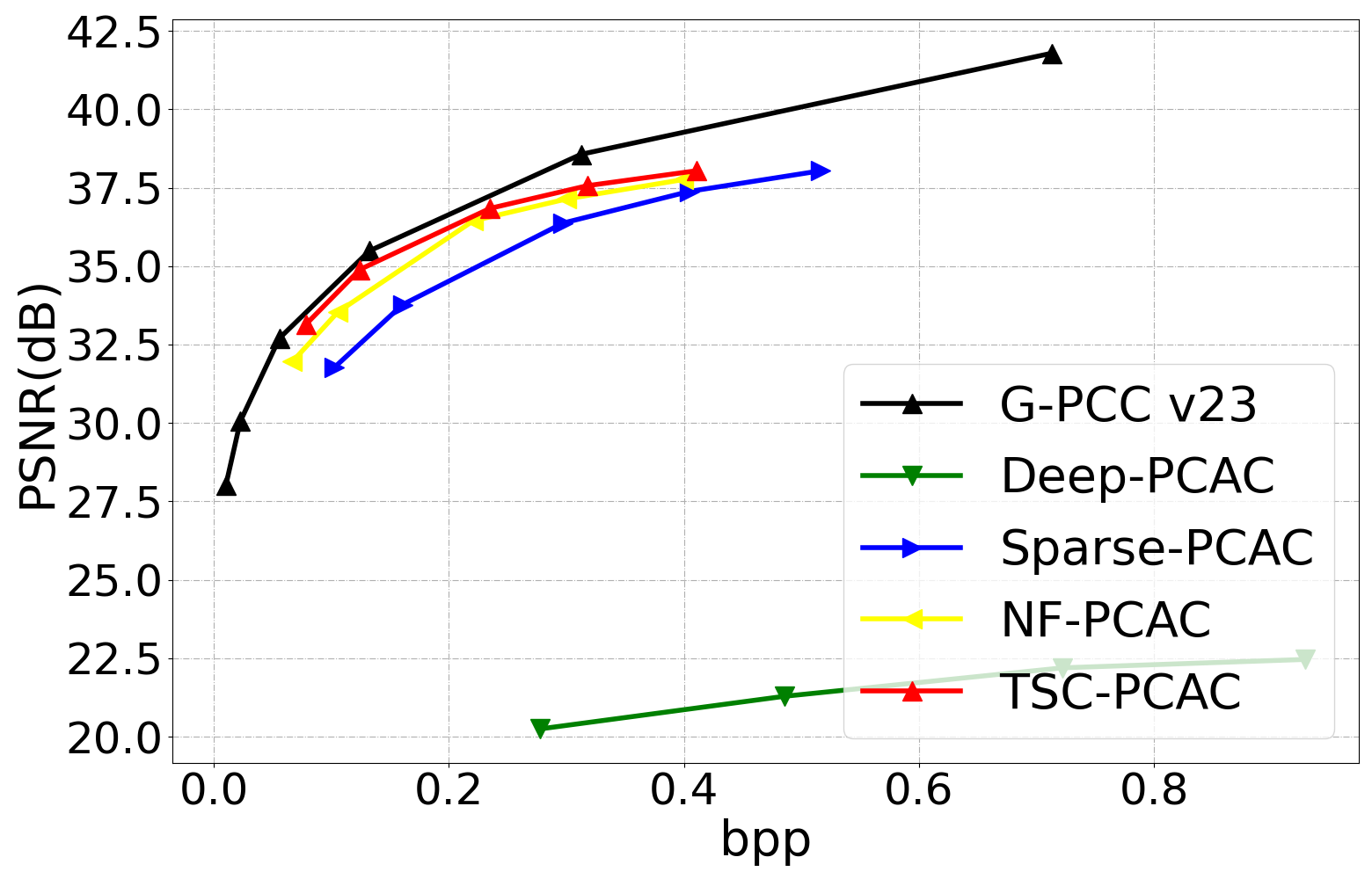}
	}
	\subfigure[] {
		\label{fig5h}
		\includegraphics[width=0.22\linewidth]{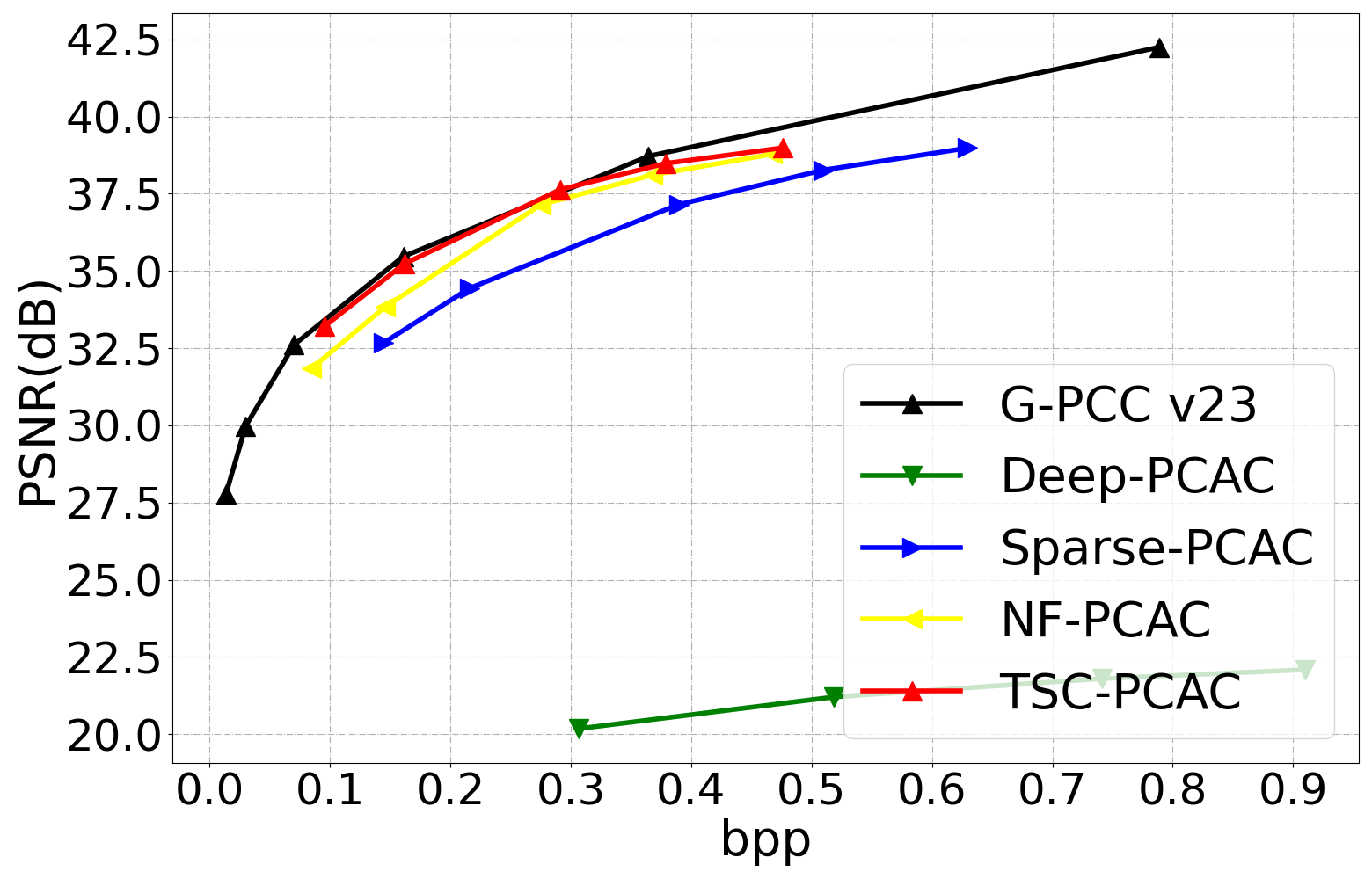}
	}
	\subfigure[] {
		\label{fig5i}
		\includegraphics[width=0.22\linewidth]{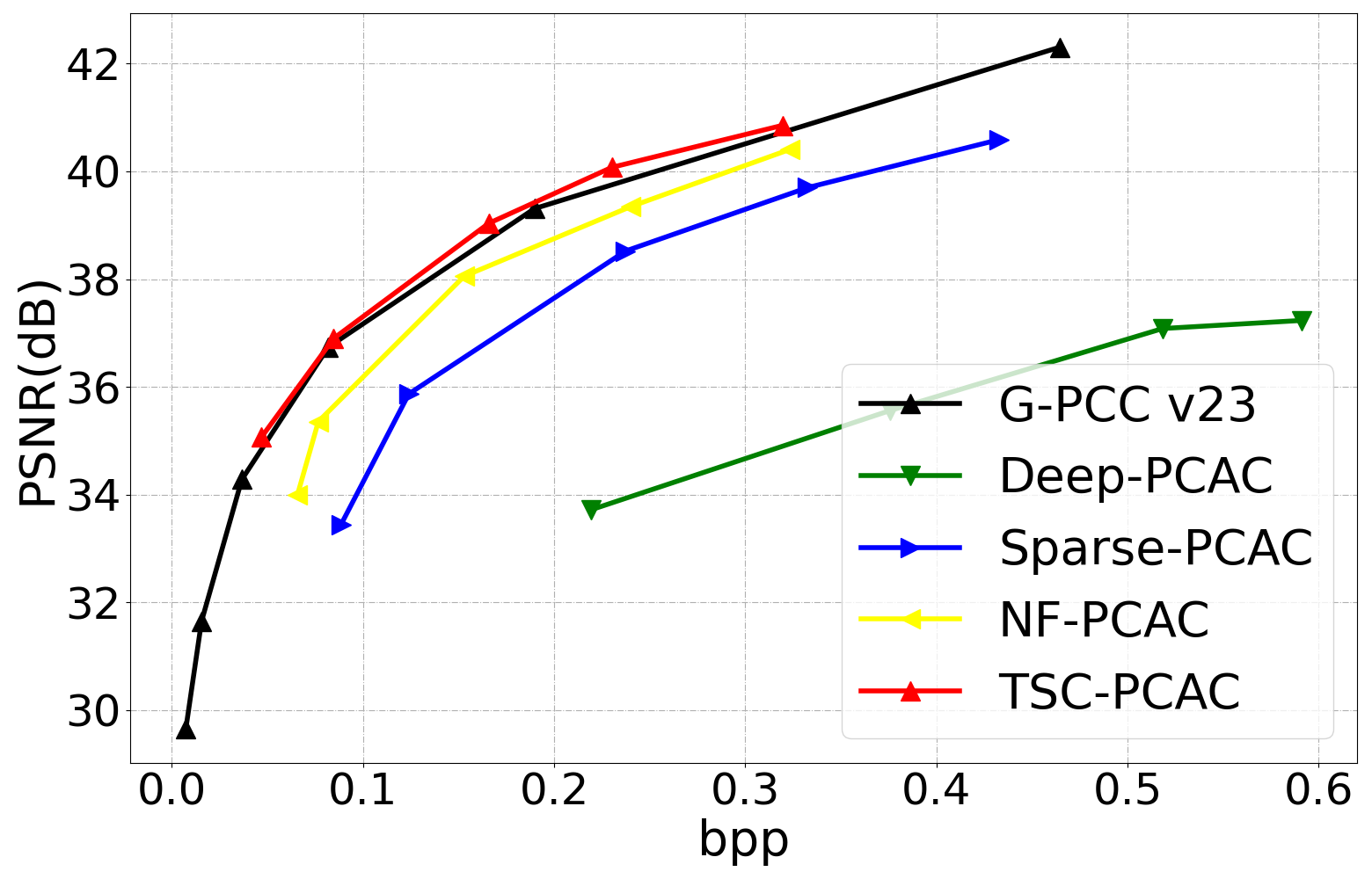}
	}
	
	\subfigure[] {
		\label{fig5j}
		\includegraphics[width=0.22\linewidth]{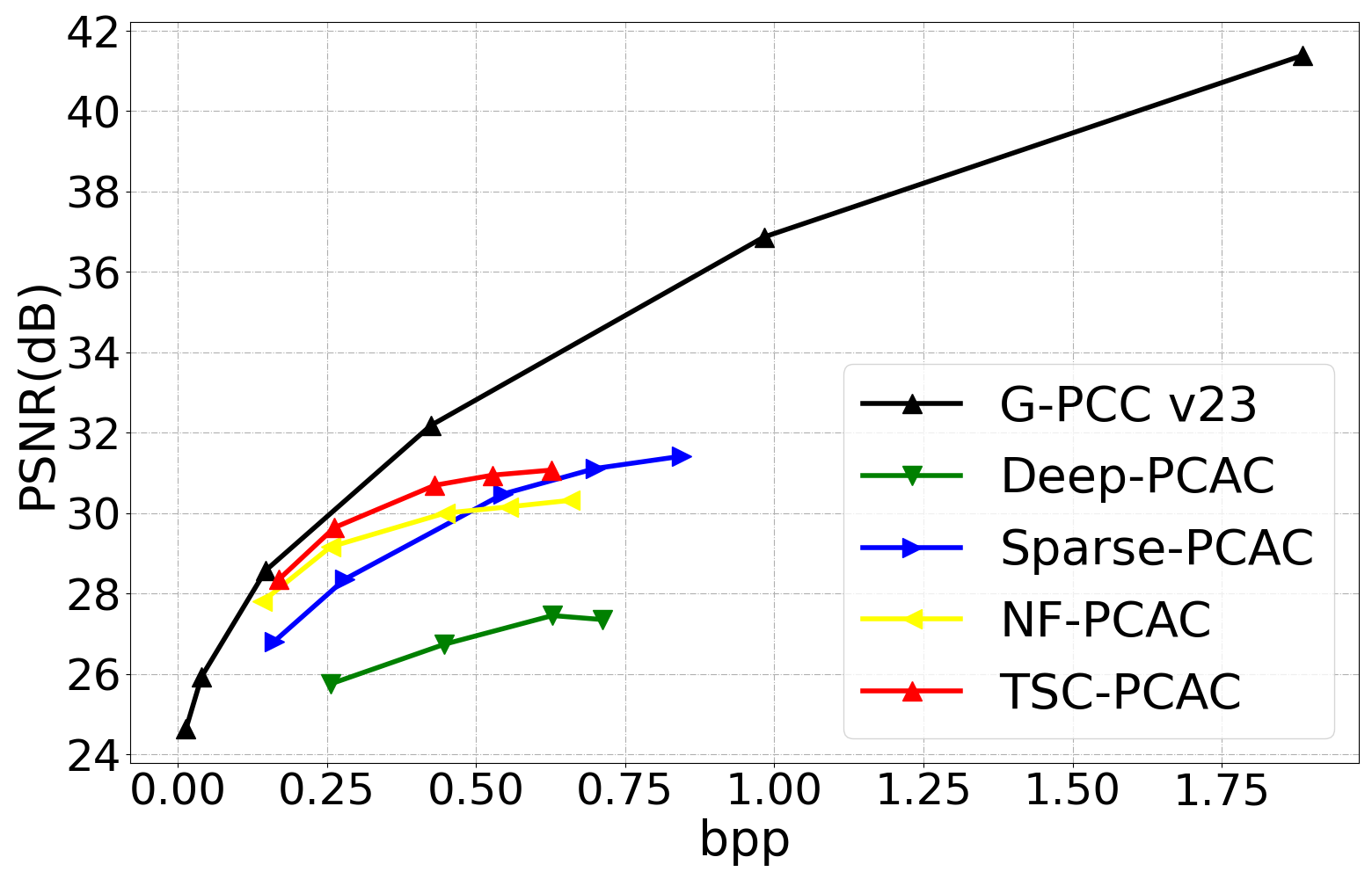}
	}
	\subfigure[] {
		\label{fig5k}
		\includegraphics[width=0.22\linewidth]{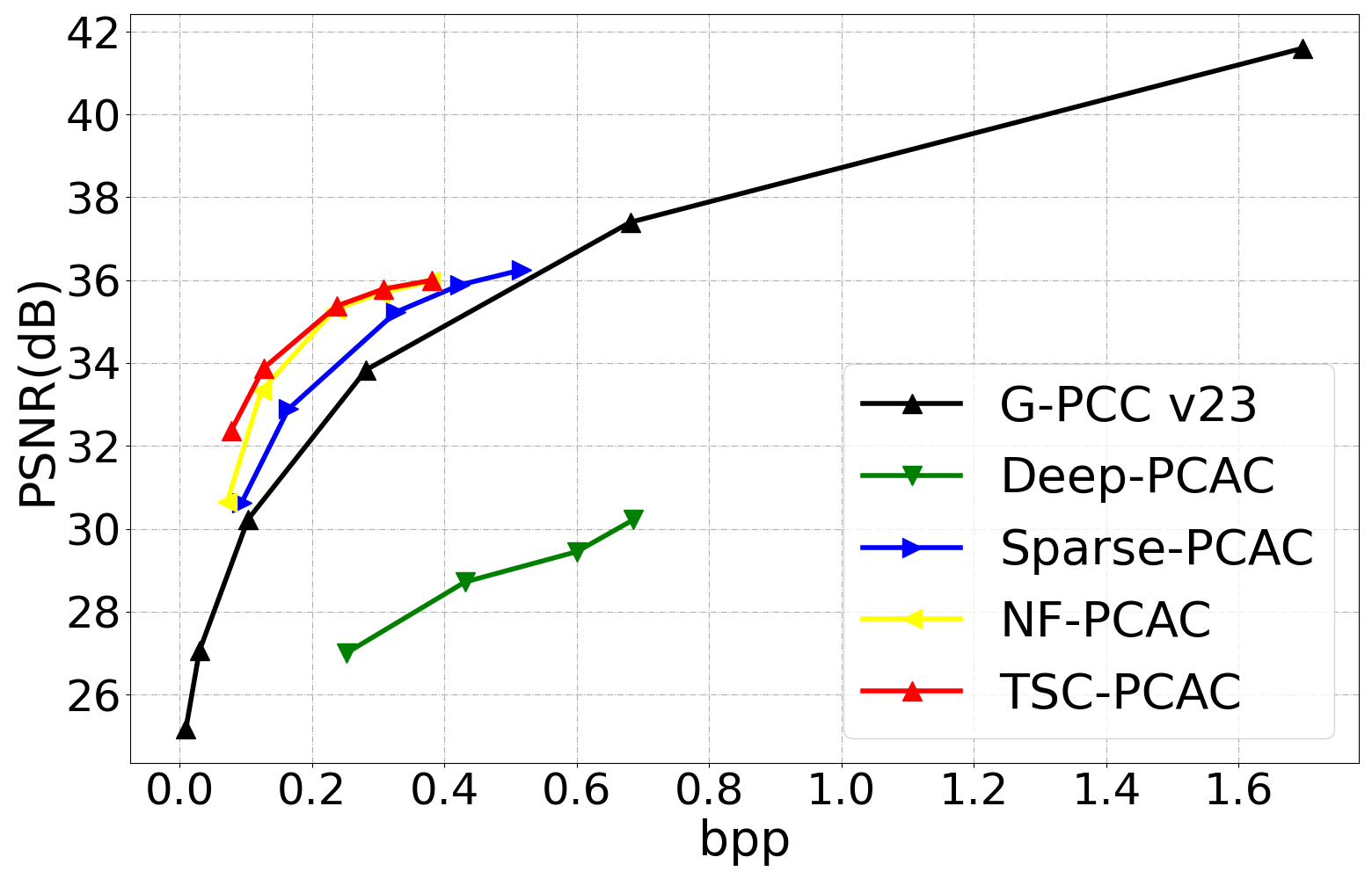}
	}
	\subfigure[] {
		\label{fig5l}
		\includegraphics[width=0.22\linewidth]{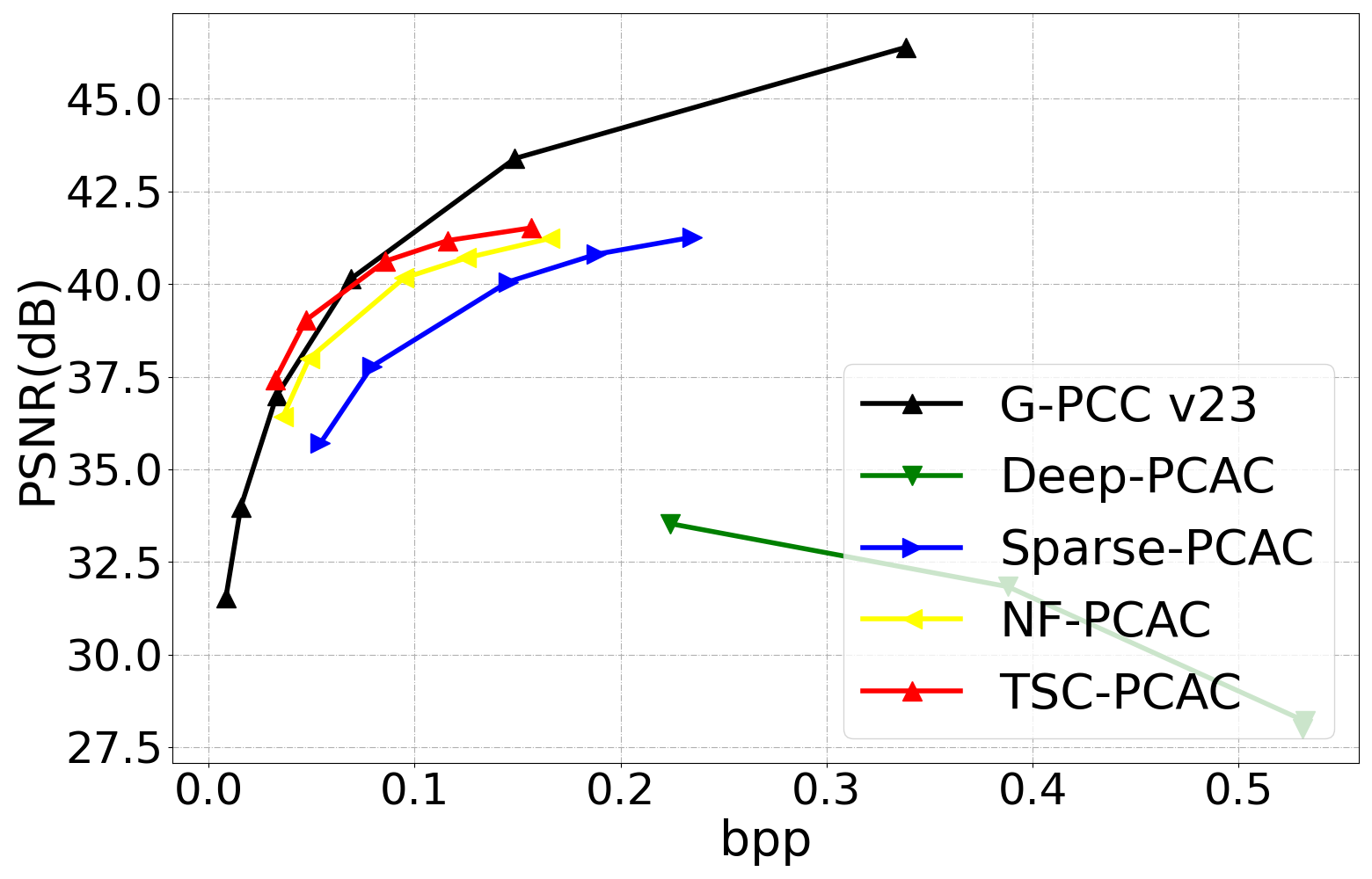}
	}
	\subfigure[] {
		\label{fig5m}
		\includegraphics[width=0.22\linewidth]{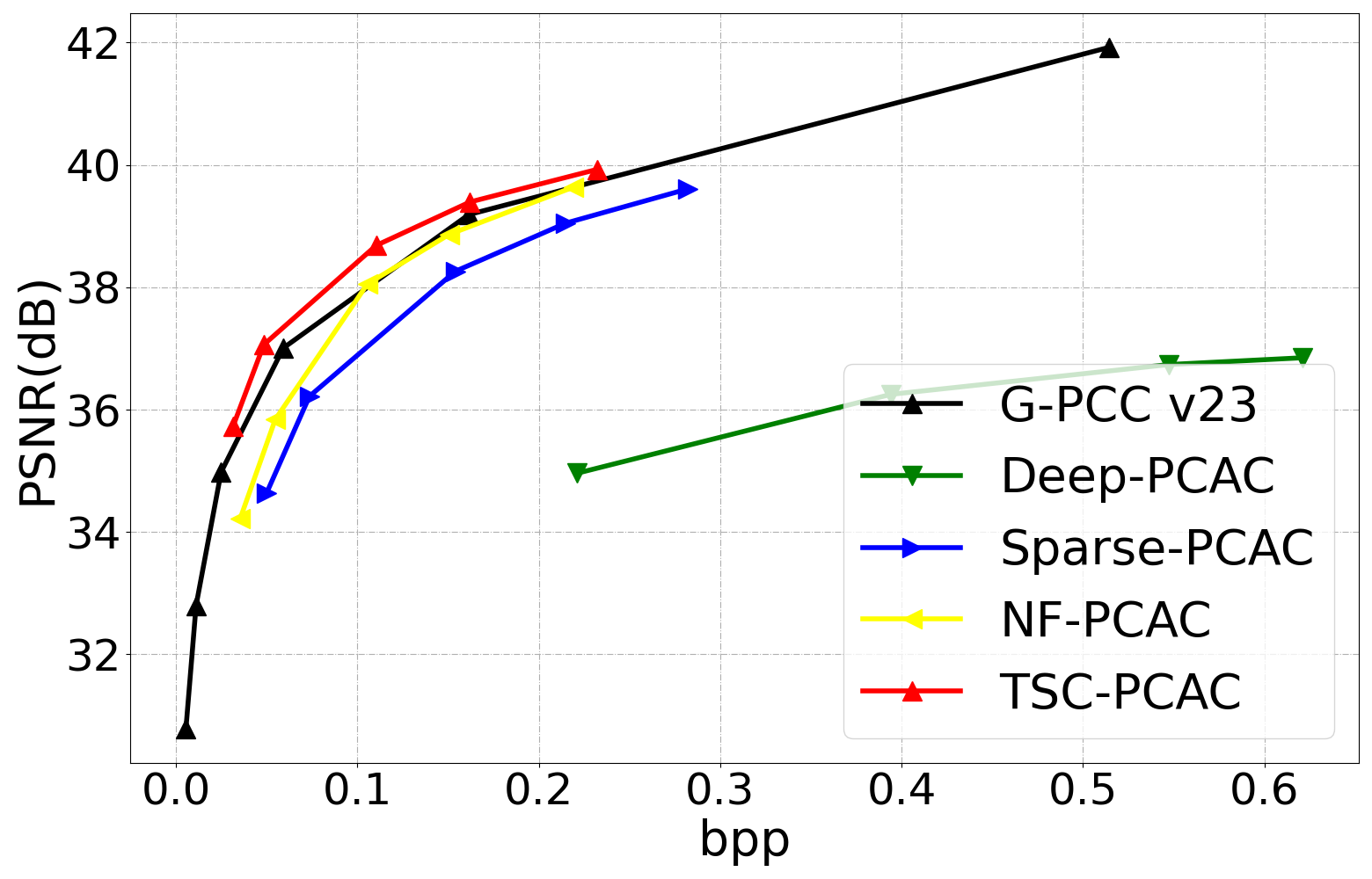}
	}
	
	\subfigure[] {
		\label{fig5n}
		\includegraphics[width=0.22\linewidth]{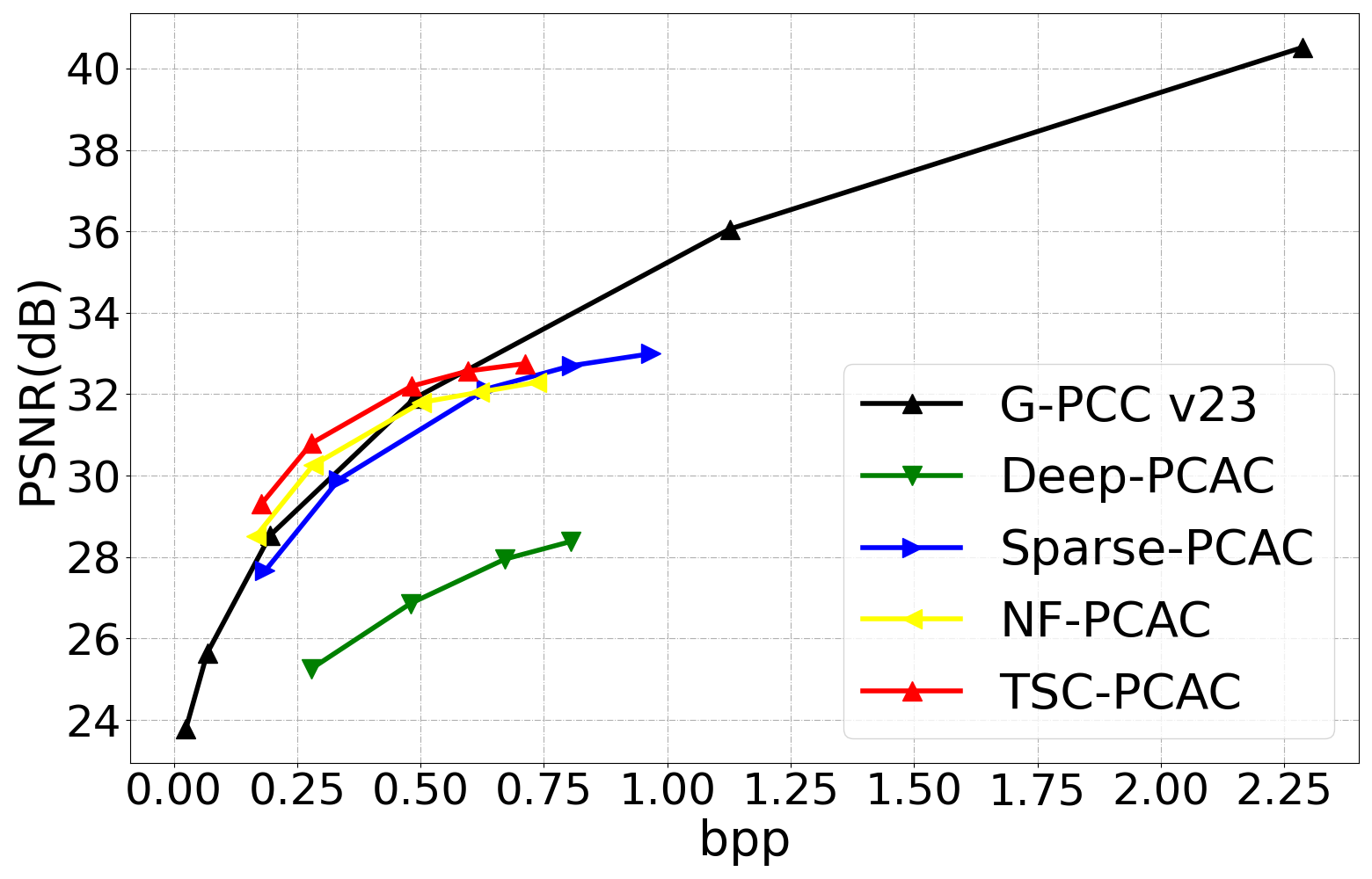}
	}
	\subfigure[] {
		\label{fig5o}
		\includegraphics[width=0.22\linewidth]{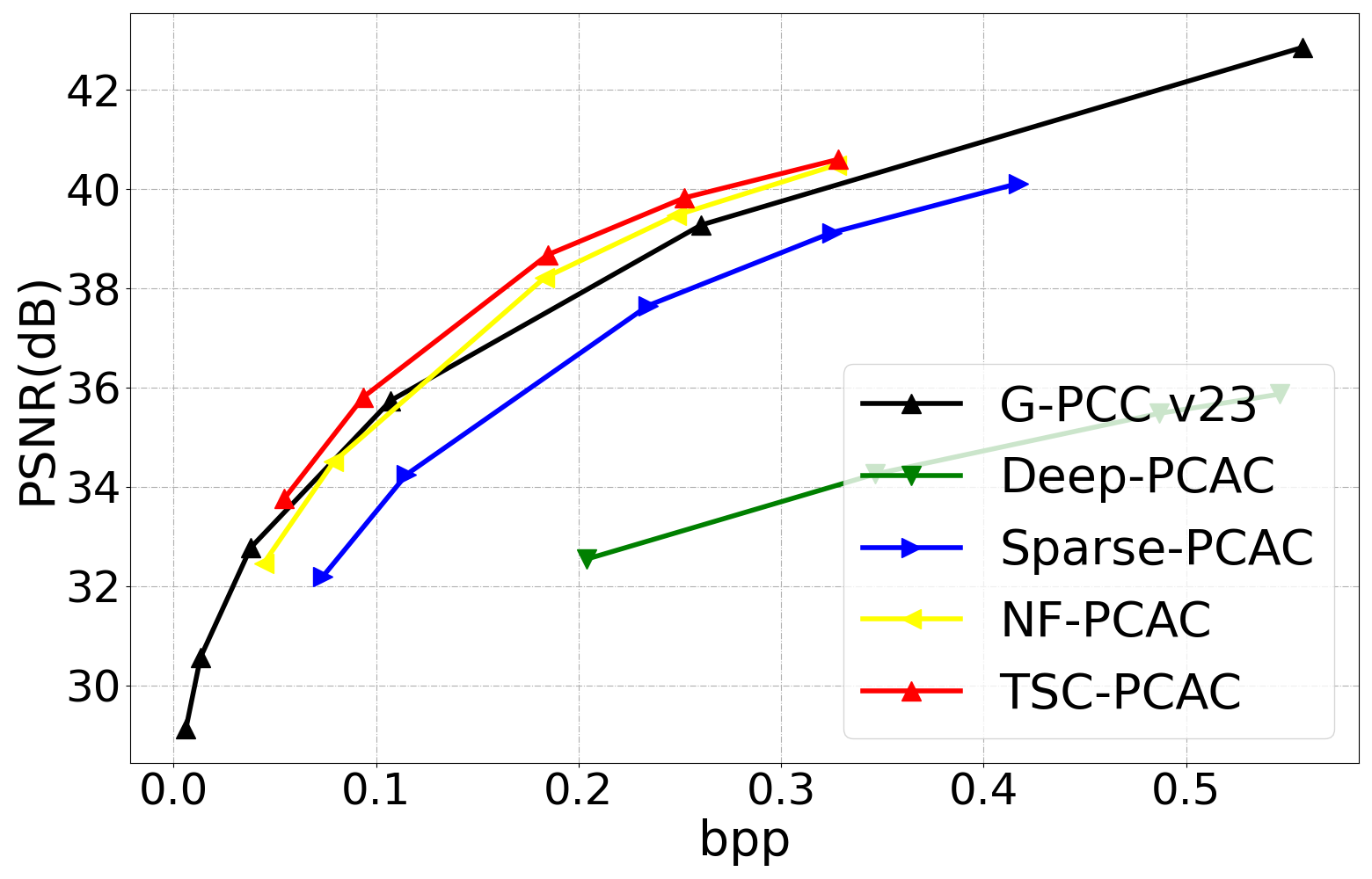}
	}
	\subfigure[] {
		\label{fig5p}
		\includegraphics[width=0.22\linewidth]{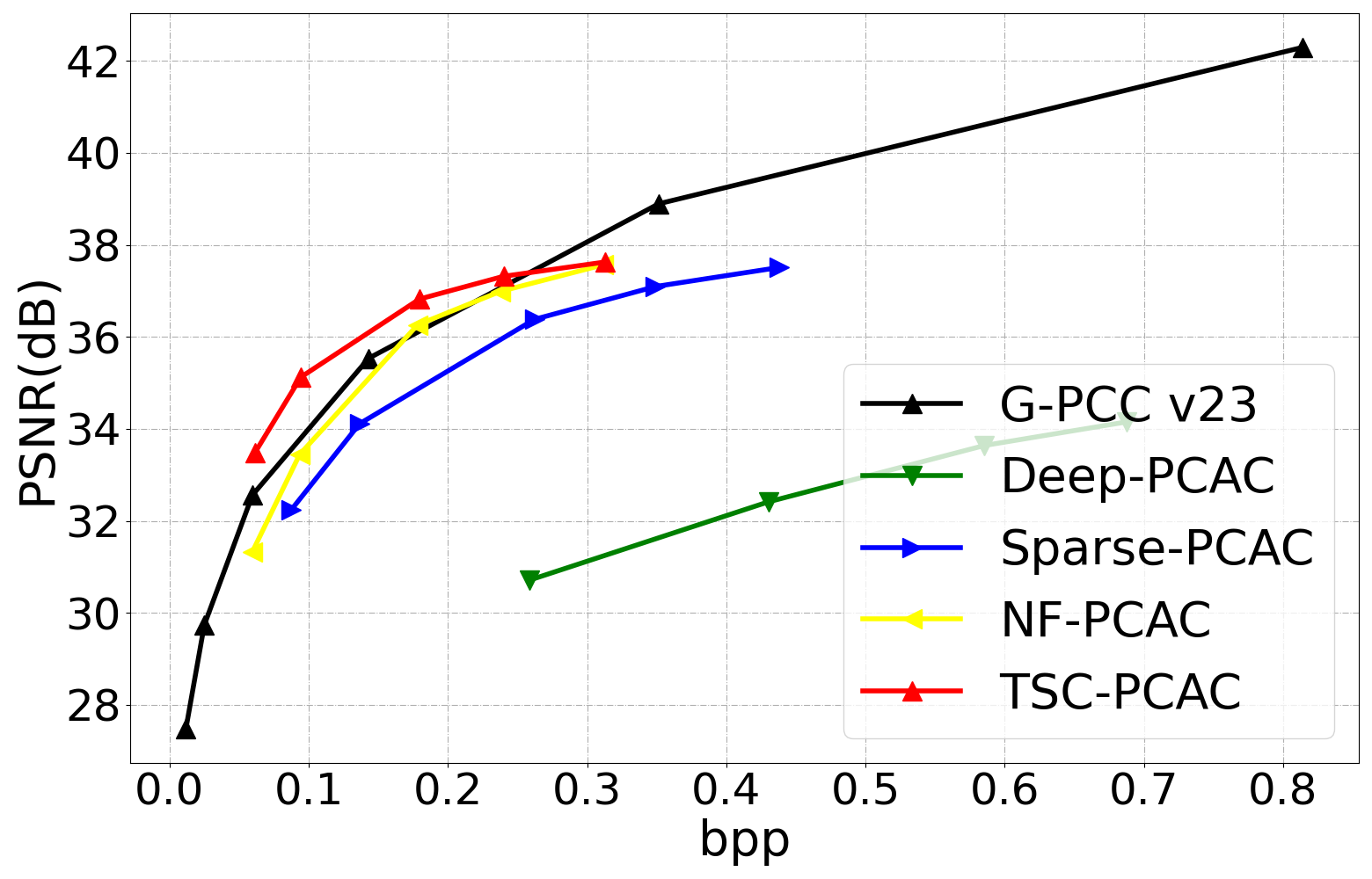}
	}
	\subfigure[] {
		\label{fig5f}
		\includegraphics[width=0.22\linewidth]{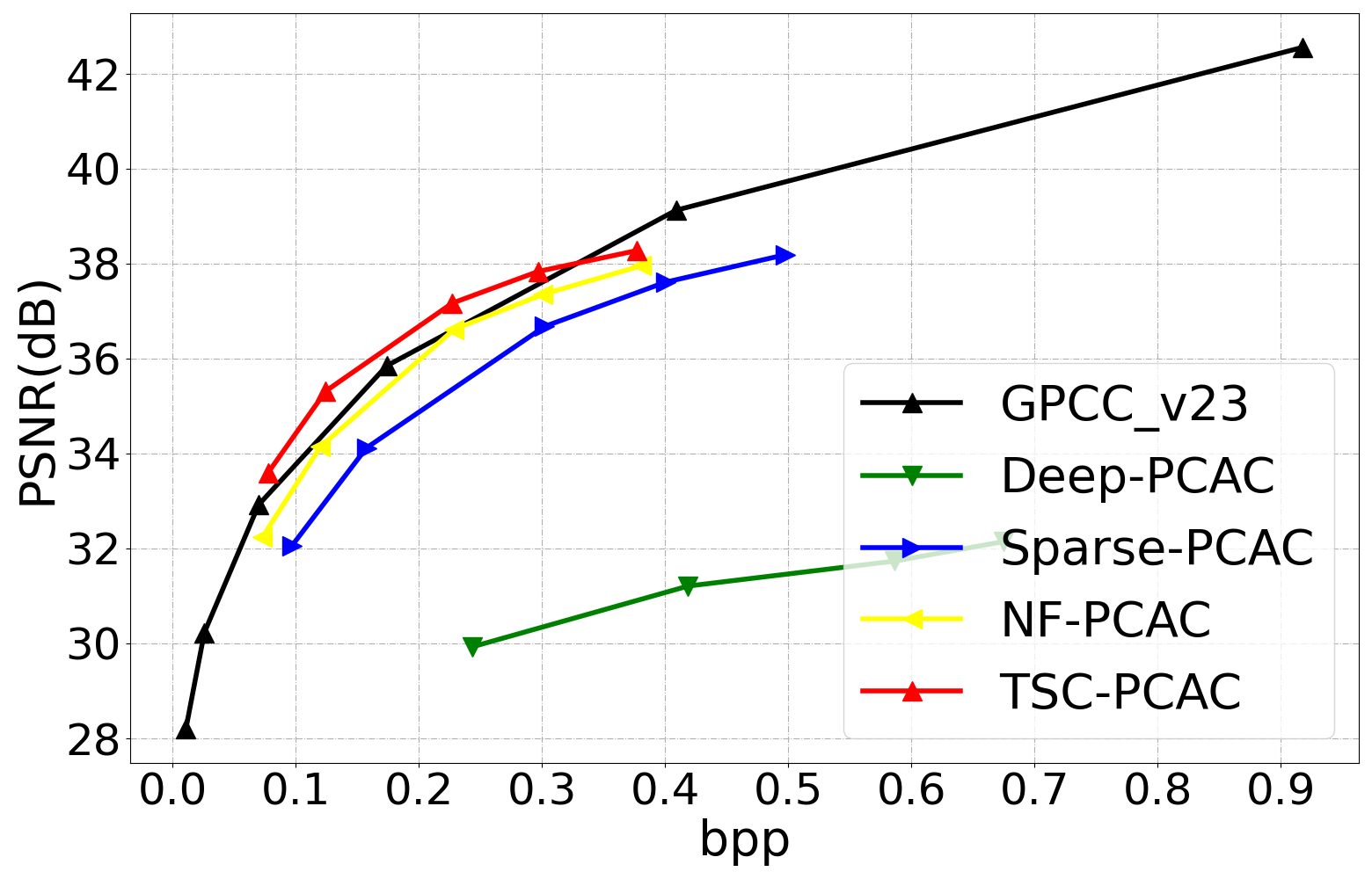}
	}
	
	\caption{RD curves of the TSC-PCAC, Sparse-PCAC, Deep-PCAC, NF-PCAC and G-PCC for different point clouds. (a) Dancer, (b) Basketball Player, (c) Redandblack, (d) Soldier, (e) Phil, (f) Rafa, (g) Sir Frederick, (h) Boxer, (i) Thaidancer, (j) Andrew*, (k) Sarah*, (l) Exercise*, (m) Longdress*, (n) Loot*, (o) Model*, (p) Average*.}
	\label{fig5}
\end{figure*}
\begin{table*}[]
	\caption{BD-BR and BD-PSNR of TSC-PCAC against the state-of-the-art. }
	\label{tab1}
	\centering
	\setlength{\tabcolsep}{1mm}
	\begin{tabular}{|c|cc|cc|cc|}
		\hline
		\multirow{2}{*}{PC}                                              & \multicolumn{2}{c|}{TSC-PCAC   vs Sparse-PCAC}       & \multicolumn{2}{c|}{TSC-PCAC   vs NF-PCAC}           & \multicolumn{2}{c|}{TSC-PCAC   vs GPCC v23}           \\ \cline{2-7}
		& \multicolumn{1}{c|}{BD-BR(\%)}     & BD-PSNR(dB) & \multicolumn{1}{c|}{BD-BR(\%)}     & BD-PSNR(dB) & \multicolumn{1}{c|}{BD-BR(\%)}     & BD-PSNR(dB)  \\ \hline
		Soldier                                                          & \multicolumn{1}{c|}{-35.09}          & 1.73          & \multicolumn{1}{c|}{-24.96}          & 1.10          & \multicolumn{1}{c|}{-8.08}           & 0.22           \\ \hline
		Redandblack                                                      & \multicolumn{1}{c|}{-30.12}          & 1.17          & \multicolumn{1}{c|}{-19.13}          & 0.59          & \multicolumn{1}{c|}{22.44}           & -0.83          \\ \hline
		Dancer                                                           & \multicolumn{1}{c|}{-49.11}          & 2.35          & \multicolumn{1}{c|}{-29.61}          & 1.25          & \multicolumn{1}{c|}{-9.30}           & 0.26           \\ \hline
		\begin{tabular}[c]{@{}c@{}}Basketball      Player\end{tabular} & \multicolumn{1}{c|}{-50.14}          & 2.18          & \multicolumn{1}{c|}{-25.68}          & 1.04          & \multicolumn{1}{c|}{-3.72}           & 0.06           \\ \hline
		Phil                                                             & \multicolumn{1}{c|}{-33.65}          & 1.48          & \multicolumn{1}{c|}{-22.05}          & 0.77          & \multicolumn{1}{c|}{-4.33}           & 0.02           \\ \hline
		Boxer                                                     & \multicolumn{1}{c|}{-44.00}          & 2.01          & \multicolumn{1}{c|}{-23.97}          & 0.85          & \multicolumn{1}{c|}{-2.43}           & 0.07           \\ \hline
		Thaidancer                                                       & \multicolumn{1}{c|}{-32.67}          & 1.15          & \multicolumn{1}{c|}{-19.54}          & 0.54          & \multicolumn{1}{c|}{28.01}           & -1.28          \\ \hline
		Rafa                                                             & \multicolumn{1}{c|}{-35.93}          & 1.58          & \multicolumn{1}{c|}{-12.85}          & 0.47          & \multicolumn{1}{c|}{18.75}           & -0.67          \\ \hline
		Sir Frederick                                                    & \multicolumn{1}{c|}{-35.47}          & 1.78          & \multicolumn{1}{c|}{-13.22}          & 0.61          & \multicolumn{1}{c|}{7.58}            & -0.31          \\ \hline
		Andrew*                                                          & \multicolumn{1}{c|}{-37.26}          & 1.44          & \multicolumn{1}{c|}{-9.44}           & 0.34          & \multicolumn{1}{c|}{-53.07}          & 2.46           \\ \hline
		David*                                                           & \multicolumn{1}{c|}{-50.57}          & 2.57          & \multicolumn{1}{c|}{-21.52}          & 0.82          & \multicolumn{1}{c|}{-0.63}           & 0.01           \\ \hline
		Exerciese*                                                       & \multicolumn{1}{c|}{-45.76}          & 1.40          & \multicolumn{1}{c|}{-30.18}          & 0.94          & \multicolumn{1}{c|}{-14.63}          & 0.34           \\ \hline
		Longdress*                                                       & \multicolumn{1}{c|}{-30.87}          & 1.24          & \multicolumn{1}{c|}{-18.38}          & 0.55          & \multicolumn{1}{c|}{-21.43}          & 0.62           \\ \hline
		Loot*                                                            & \multicolumn{1}{c|}{-39.88}          & 2.19          & \multicolumn{1}{c|}{-11.24}          & 0.48          & \multicolumn{1}{c|}{-14.39}          & 0.58           \\ \hline
		Model*                                                           & \multicolumn{1}{c|}{-44.87}          & 1.77          & \multicolumn{1}{c|}{-24.91}          & 1.02          & \multicolumn{1}{c|}{-18.72}          & 0.54           \\ \hline
		Sarah*                                                           & \multicolumn{1}{c|}{-54.00}          & 2.39          & \multicolumn{1}{c|}{-26.13}          & 0.86          & \multicolumn{1}{c|}{0.12}            & -0.30          \\ \hline
		\textbf{Average}                                                 & \multicolumn{1}{c|}{\textbf{-36.62}} & \textbf{1.66} & \multicolumn{1}{c|}{\textbf{-22.27}} & \textbf{0.85} & \multicolumn{1}{c|}{\textbf{1.26}}   & \textbf{-0.11} \\ \hline
		\textbf{Average*}                                                & \multicolumn{1}{c|}{\textbf{-38.53}} & \textbf{1.72} & \multicolumn{1}{c|}{\textbf{-21.30}} & \textbf{0.80} & \multicolumn{1}{c|}{\textbf{-11.19}} & \textbf{0.34}  \\ \hline
	\end{tabular}
\end{table*}
	\subsubsection{Coding Methods} Four point cloud compression methods were compared with the proposed TSC-PCAC, including Deep-PCAC \cite{7}, Sparse-PCAC \cite{8}, {NF-PCAC} \cite{17} and G-PCC v23. For a fair comparison, we retrained the Sparse-PCAC and NF-PCAC models according to the training conditions of TSC-PCAC. As for the Deep-PCAC model, since the authors have provided pre-trained models, we used them to obtain coding results. The G-PCC encoding settings were configured with default settings for dense point clouds and {the attribute encoding mode was RAHT. Quantization Parameters (QP) followed the default settings of Common Test Conditions (CTCs) \cite{testcondition}}.

	\subsubsection{Quality Metrics} Peak Signal-to-Noise Ratio (PSNR) was used as the quality metric for distortion measurement \cite{33}, and the rate was measured in bit per point ($bpp$). We also used Bjøntegaard Delta bitrate (BD-BR) to evaluate the coding performance. Additionally, we used Bjøntegaard Delta PSNR (BD-PSNR) to compare the reconstruction quality of different codecs\cite{bjontegaard2001calculation}. It's worth noting that the PSNR of attribute in the paper is calculated from the luminance component and the geometry {is losslessly} coded.
	
\subsubsection{Databases} The dataset utilized consists of 8iVFB \cite{30}, Owlii \cite{31}, 8iVSLF \cite{krivokuca20188i}, Volograms \cite{pages2021volograms}, and MVUB \cite{32}. 	
	{
 We selected the \textit{Longdress, Loot, Exercise, Model, Andrew, Sarah, and David} point cloud sequences for training, and selected \textit{Redandblack, Soldier, Basketball Player, Dancer, Thaidancer, Boxer, Rafa, Sir Frederick and Phil} point cloud sequences for testing.
 \textit{Longdress, Loot, Soldier}, and \textit{Redandblack} are from 8iVFB. \textit{Andrew, Sarah, David}, and \textit{Phil} are from MVUB. \textit{Exercise, Model, Basketball Player} and \textit{Dancer} are from Owlii. \textit{Rafa} and \textit{Sir Frederick} are from Volograms. \textit{Thaidancer} and \textit{Boxer} are from 8iVSLF.
   For \textit{Boxer} and \textit{Thaidancer}, following \cite{nguyen2021lossless}, we quantified them to 10-bit. For \textit{Rafa} and \textit{Sir Frederick}, we sampled them from mesh data to dense colored point cloud sequences with a precision of 10-bit. For the sequences of MVUB, we used point cloud sequences with a precision of 10-bit.}

	\subsubsection{Training Settings} To reduce the data volume and accelerate training, the point clouds are resampled.
	{Specifically,  $K$ cluster centers of a point cloud with $N$ points were obtained by farthest point sampling, where $K=Ceil\left(\frac{N}{100000}\right)$.
	Then, we used the k-Nearest Neighbor (kNN) algorithm to cluster the surrounding 100,000 points on cluster centers. Resampling was applied to the first 100 frames of each point cloud sequence in the training set. The point clouds in the testing set were not resampled.} {To further validate the compression performance, we included more test sequences where the previous frames are used as training and the rest subsequent frames are used as testing data. Their coding results were marked with `*'.} We set $\lambda_A$ as 400, 1000, 4000, 8000, and 16000 to obtain different bitrate points. For efficient training, we first trained a model with  $\lambda_A$ as 16000.  {We then fine-tuned the model 50 epochs at a learning rate of $10^{-5}$ to obtain different models at different values of $\lambda_A$.} During the training, we found that it was easier to train the model without the channel context module first and then jointly train it. Considering the trade-off between coding performance and {computational} complexity, we set the number of channel divisions $C$ in the channel context module as 8. All the experiments were conducted on a workstation with an Intel Core i9-10900 CPU and an NVIDIA GeForce RTX 3090 GPU.

%

%
%
\begin{figure*}[!t]
	\centering
	\subfigure[] {
		\label{fig6a}
		\includegraphics[width=0.16\linewidth]{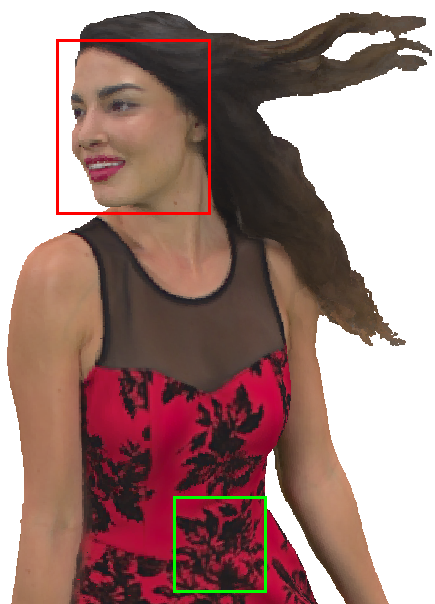}}
	\subfigure[] {
		\label{fig6b}
		\includegraphics[width=0.11\linewidth]{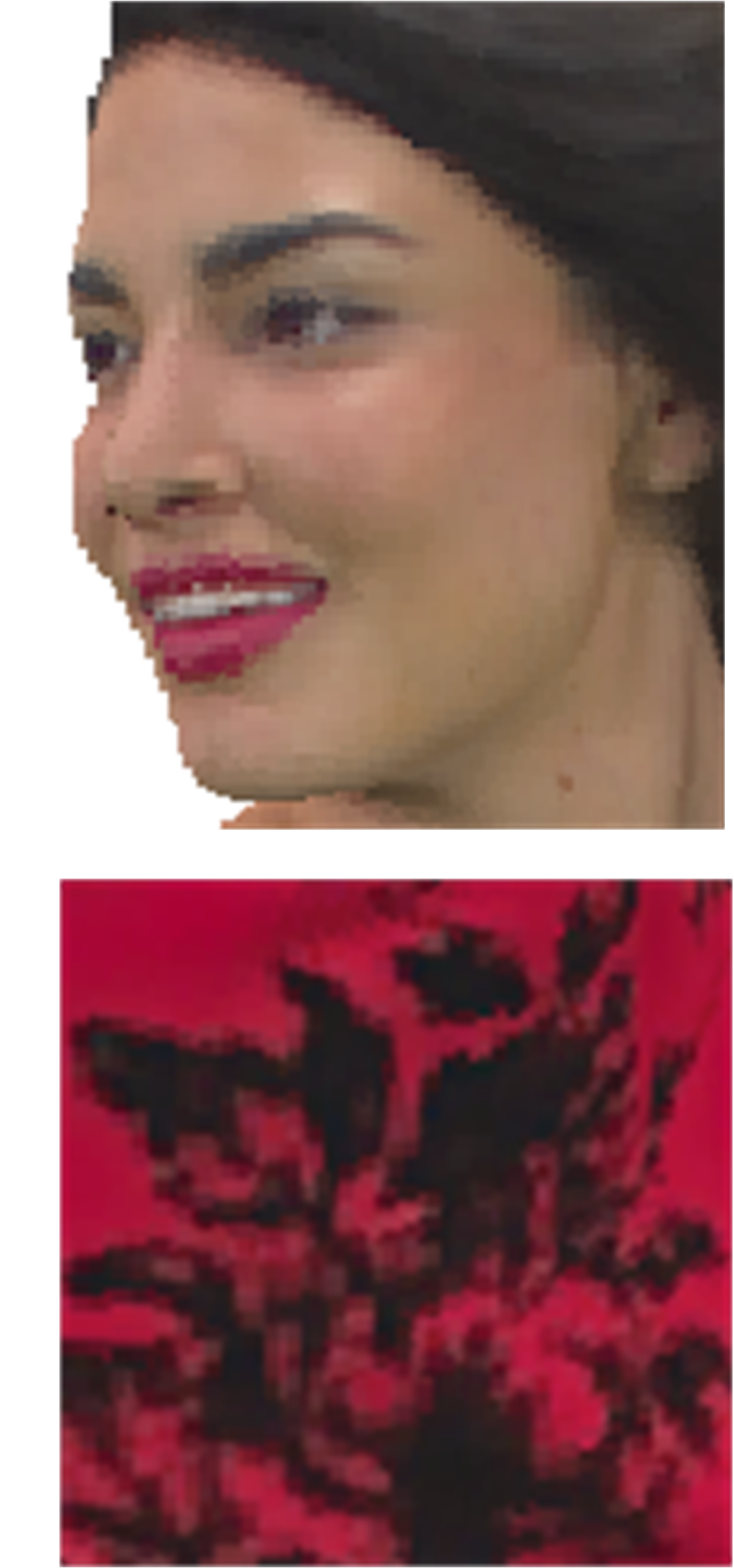}
	}
	\subfigure[] {
		\label{fig6d}
		\includegraphics[width=0.11\linewidth]{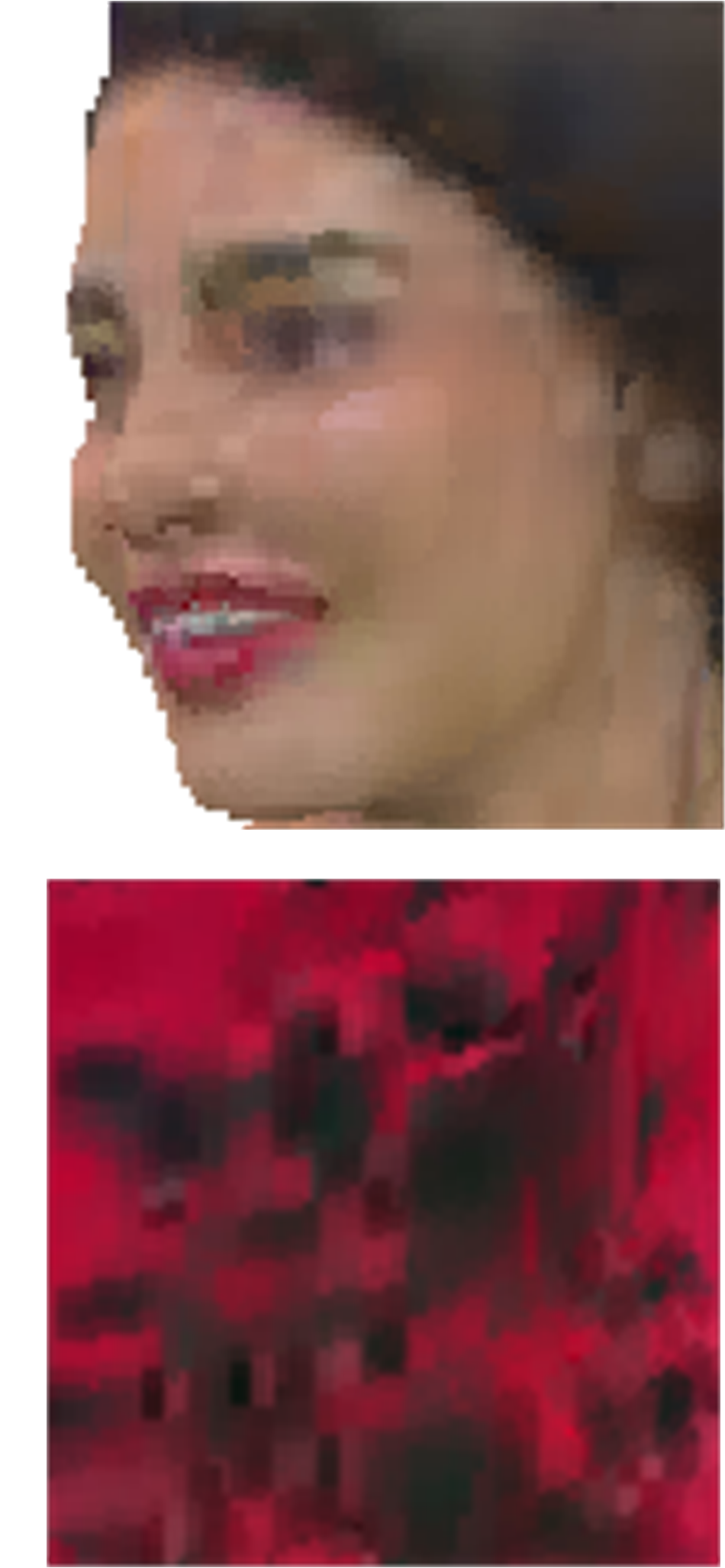}
	}
	\subfigure[] {
		\label{fig6e}
		\includegraphics[width=0.11\linewidth]{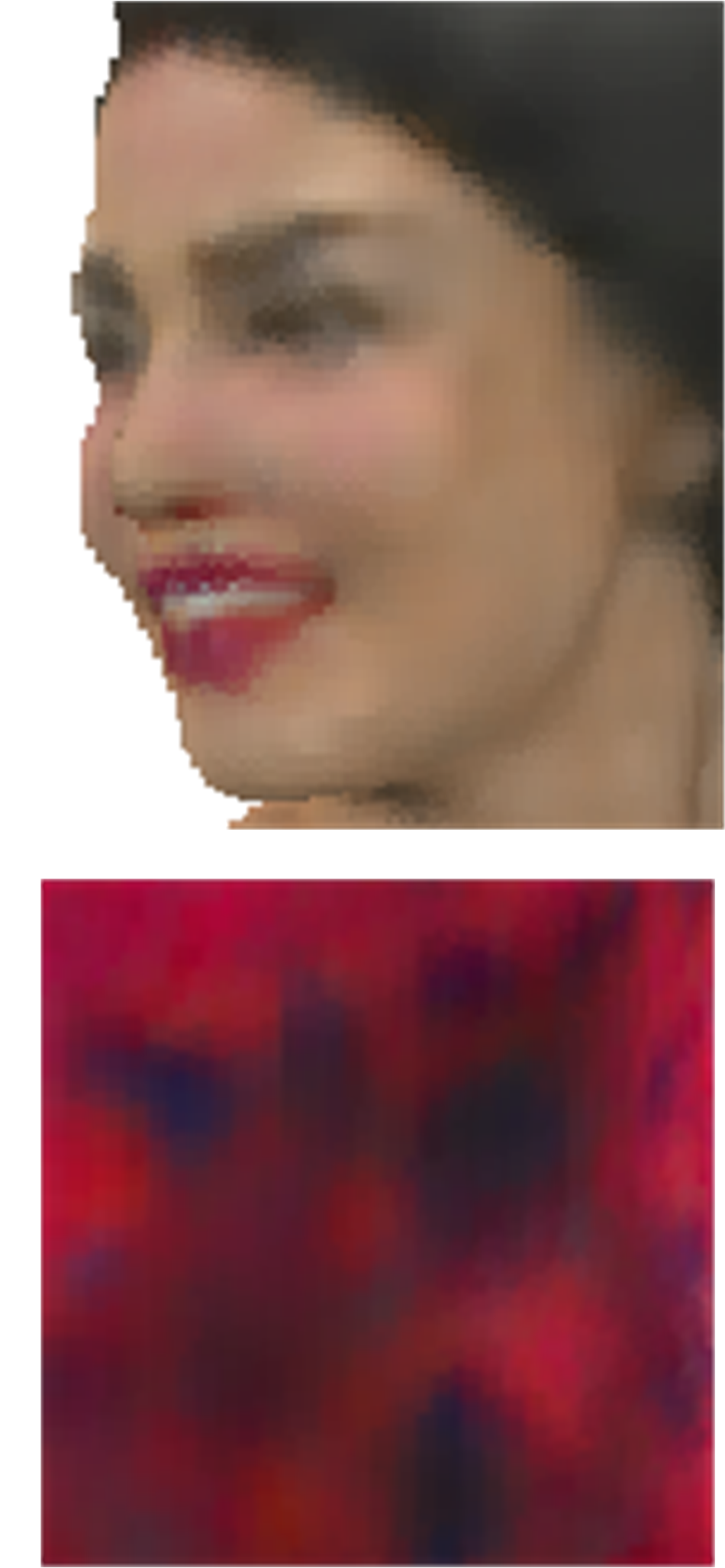}
	}
	\subfigure[] {
		\label{fig6f}
		\includegraphics[width=0.11\linewidth]{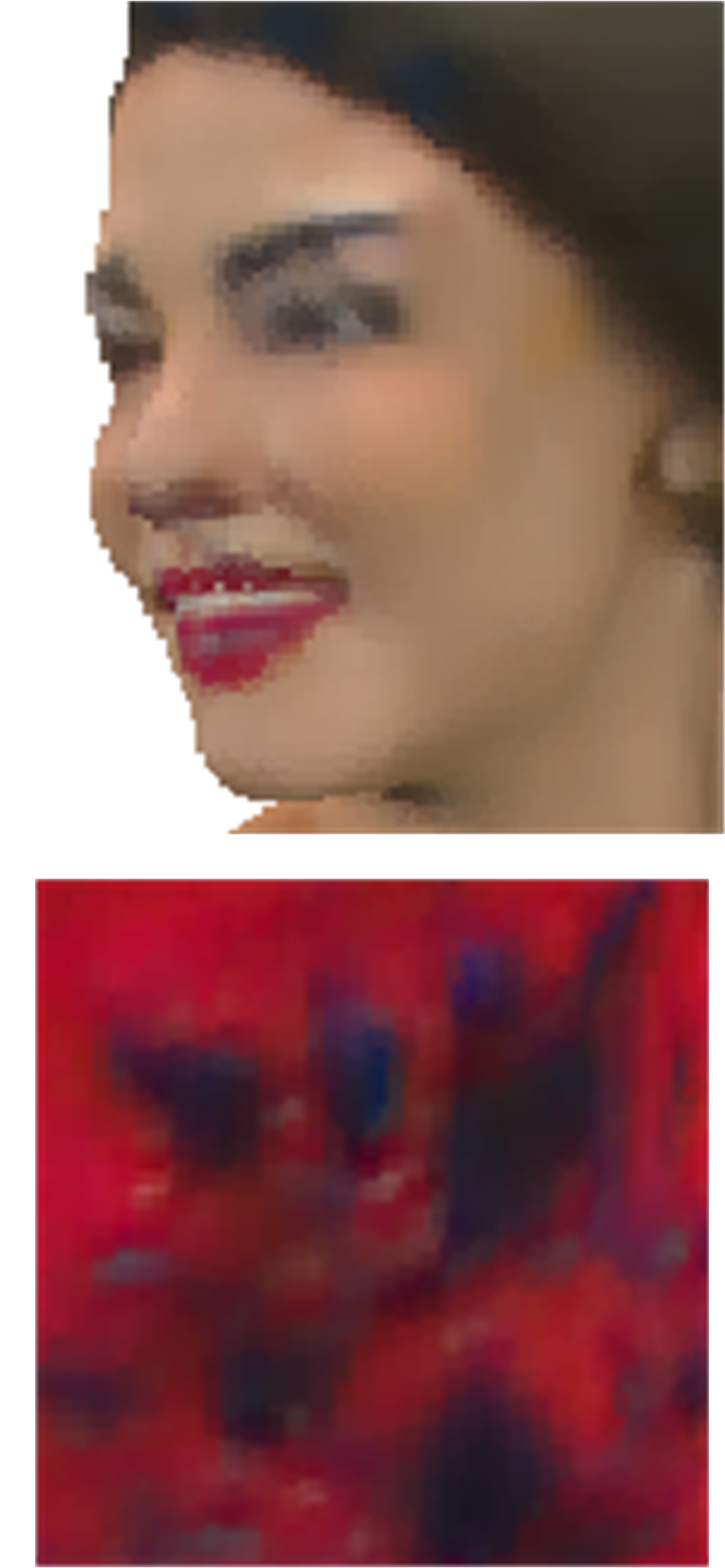}
	}
	\subfigure[] {
		\label{fig6g}
		\includegraphics[width=0.11\linewidth]{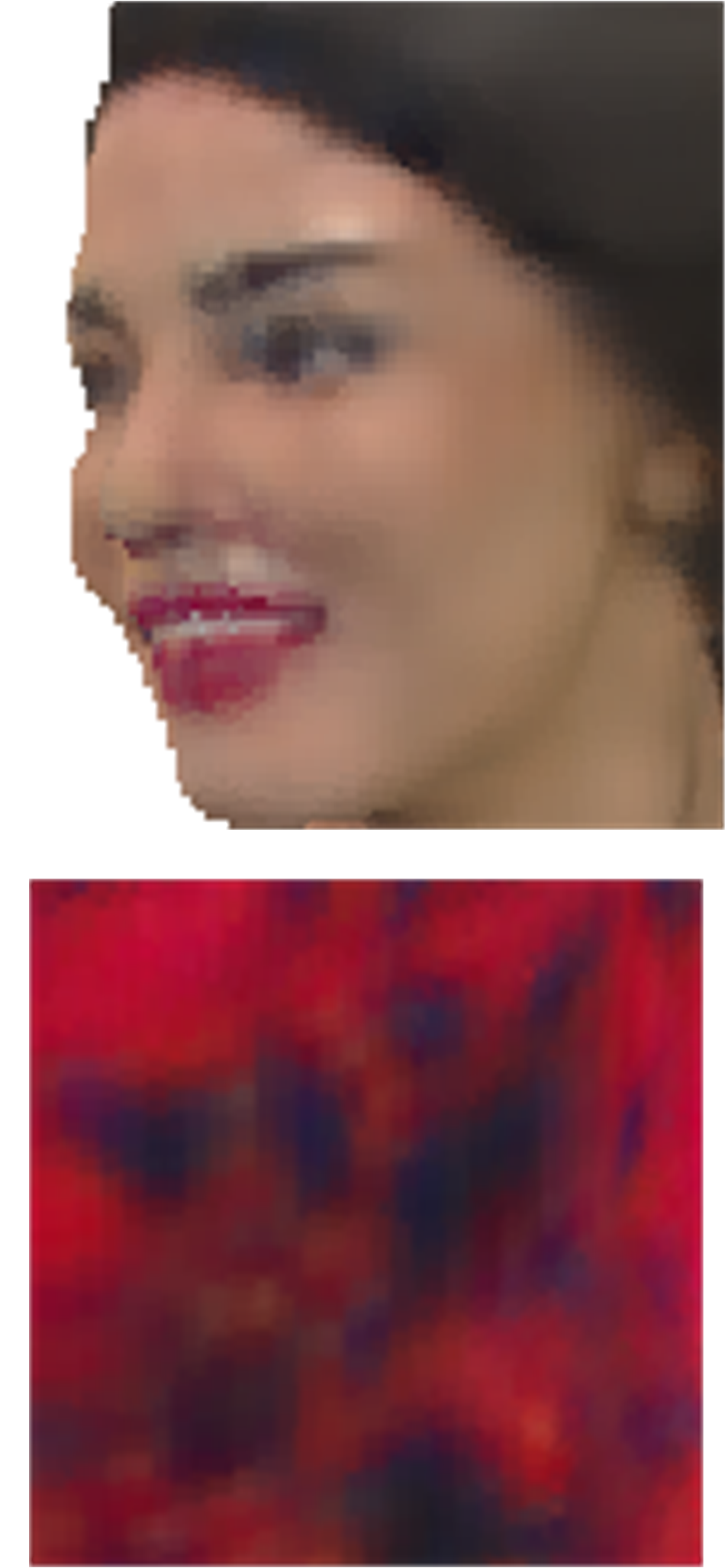}
	}
	
	\subfigure[] {
		\label{fig6so}
		\includegraphics[width=0.2\linewidth]{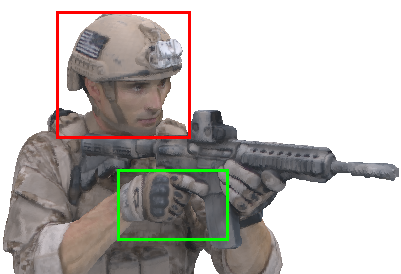}}
	\subfigure[] {
		\label{fig6h}
		\includegraphics[width=0.11\linewidth]{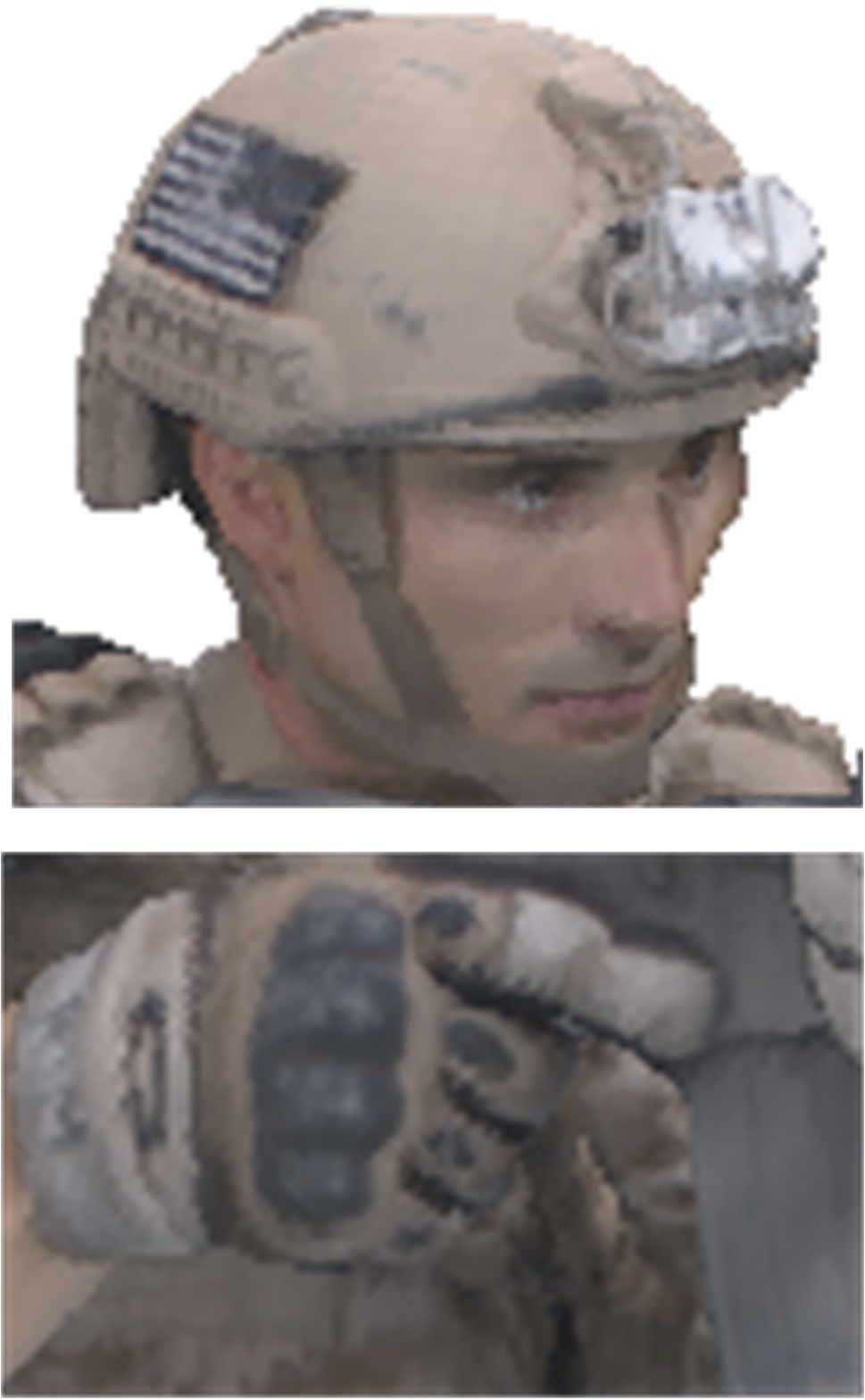}
	}
	\subfigure[] {
		\label{fig6j}
		\includegraphics[width=0.11\linewidth]{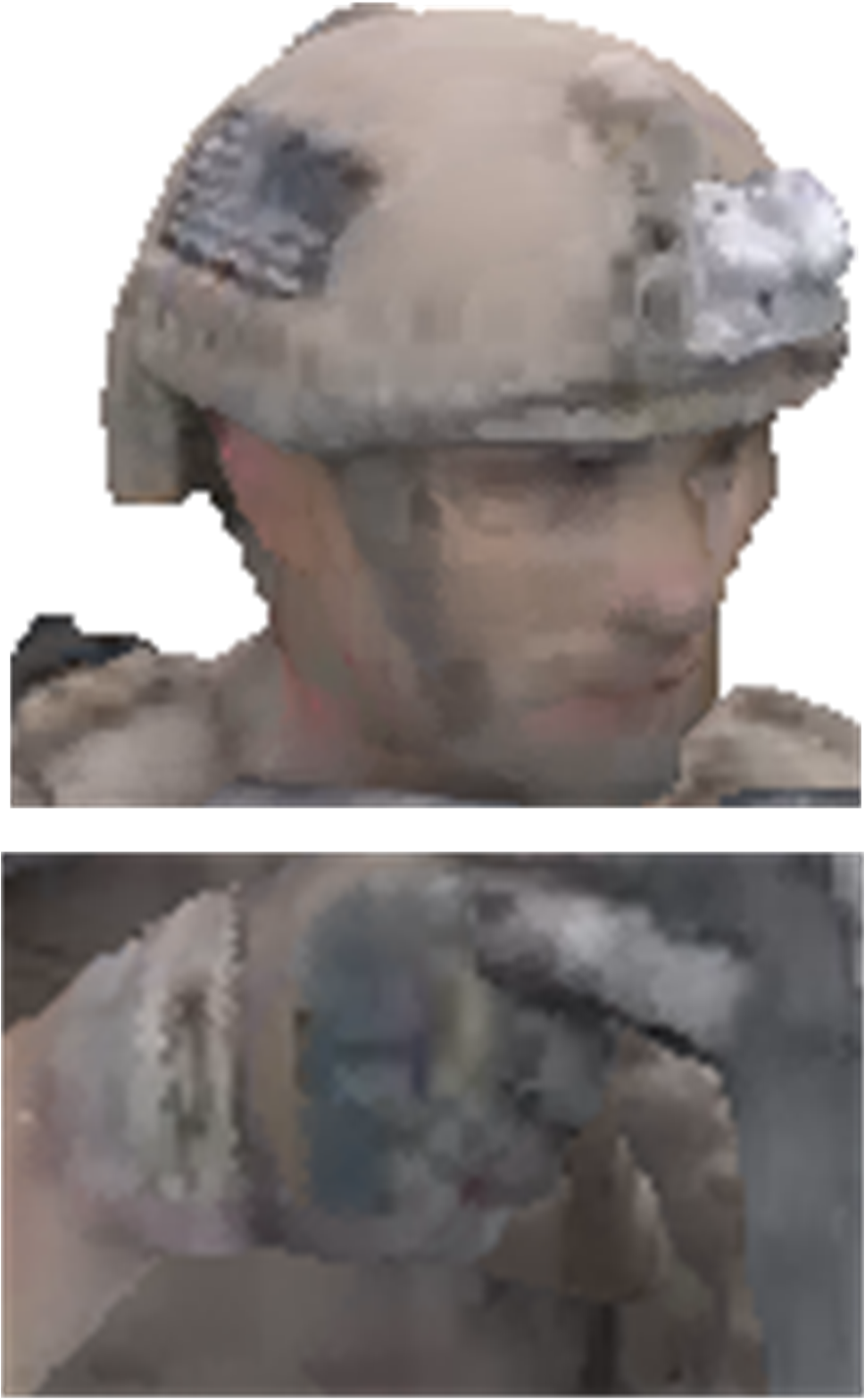}
	}
	\subfigure[] {
		\label{fig6k}
		\includegraphics[width=0.11\linewidth]{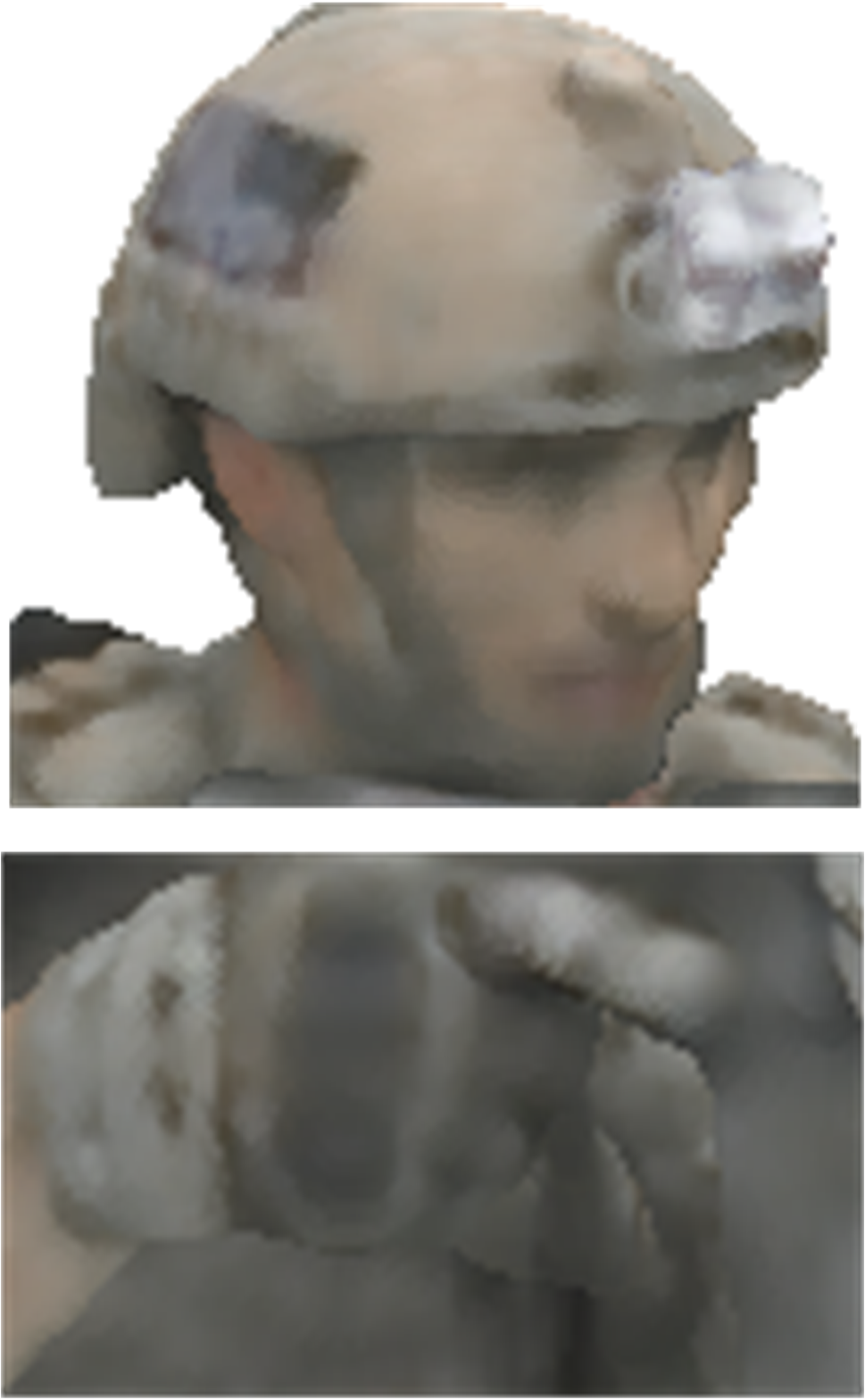}
	}
	\subfigure[] {
		\label{fig6l}
		\includegraphics[width=0.11\linewidth]{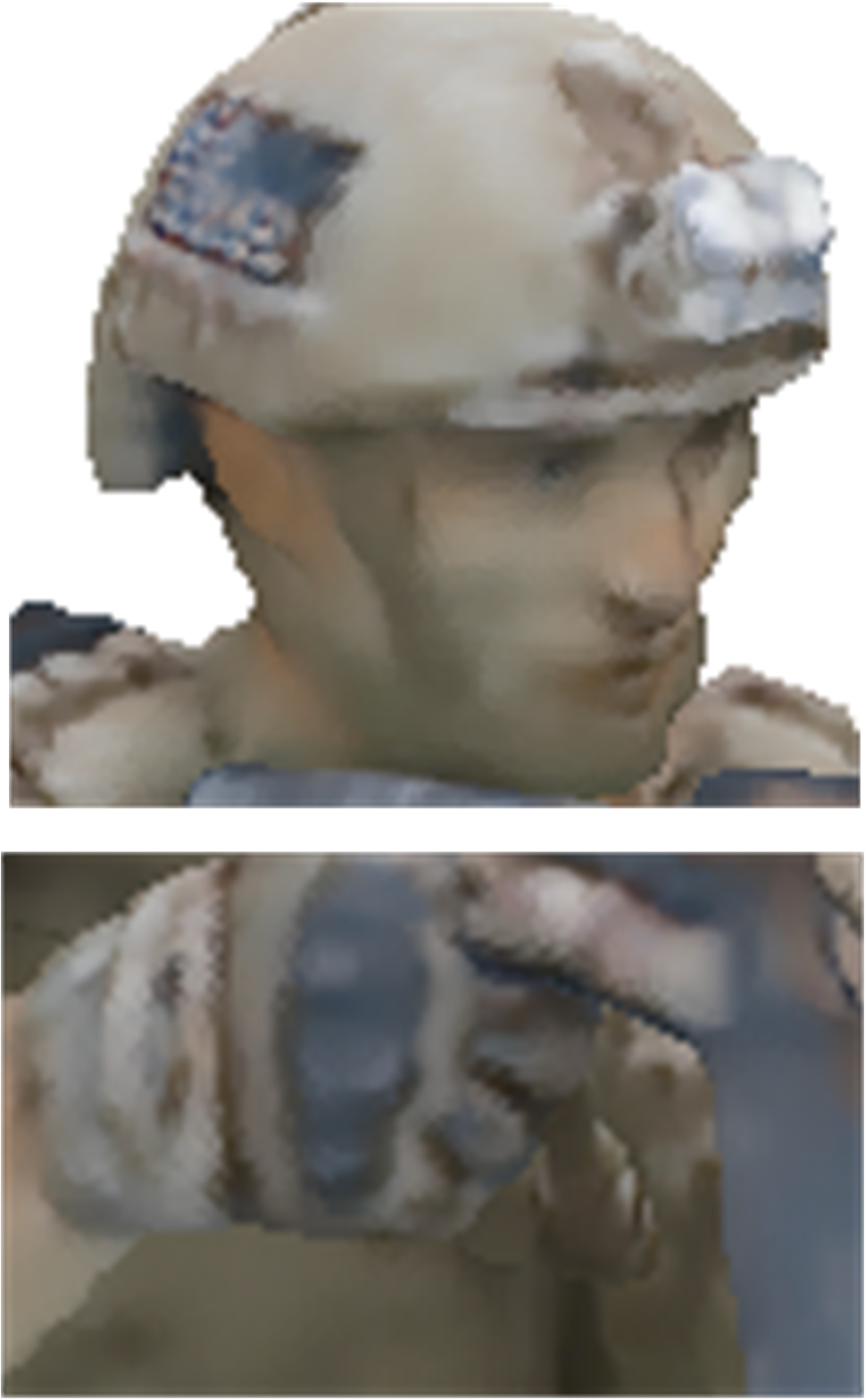}
	}
	\subfigure[] {
		\label{fig6m}
		\includegraphics[width=0.11\linewidth]{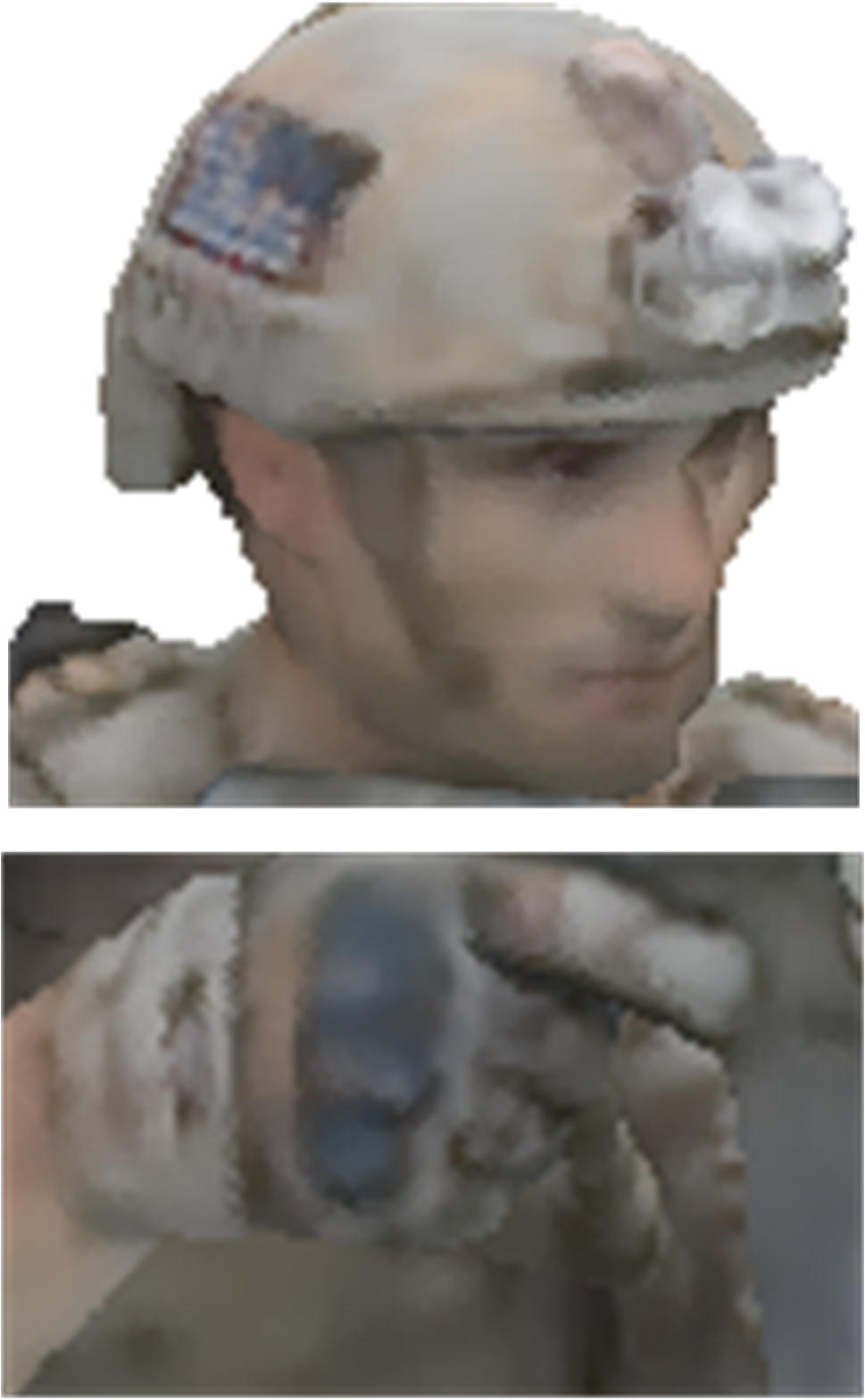}
	}
	
	\caption{{Visual quality comparison. Digits in blanks denote the bit rate and quality, i.e. (bpp, PSNR).  (a) \textit{Redandblack}, (b) Enlarged \textit{Redandblack}, (c) G-PCC v23 (0.102, 33.10dB), (d) Sparse-PCAC (0.143, 31.75dB), (e) NF-PCAC (0.116, 32.21dB), (f) TSC-PCAC (0.132, 33.24dB), (g) \textit{Soldier}, (h) Enlarged \textit{Soldier}, (i) G-PCC v23 (0.075, 30.42dB), (j) Sparse-PCAC (0.100, 29.75dB), (k) NF-PCAC (0.083, 29.87dB), (l) TSC-PCAC (0.084, 31.23dB).}
	 }
	\label{fig6}
\end{figure*}

	\subsection{Coding Efficiency Evaluation}	
	Fig. \ref{fig5} illustrates the coding performance comparison of Deep-PCAC, Sparse-PCAC, G-PCC, NF-PCAC and TSC-PCAC, where the vertical-axis is PSNR and the horizontal-axis is bit rate in terms of $bpp$. {It can be observed that the Sparse-PCAC significantly outperforms the Deep-PCAC, but it is worse than the G-PCC in the majority of point clouds.} {NF-PCAC has higher compression efficiency than Sparse-PCAC. As for the proposed TSC-PCAC, it outperforms the NF-PCAC by a large margin in all point clouds.}
	{Our TSC-PCAC achieves the best performance on most test point clouds, such as \textit{Soldier} and \textit{Dancer}. However, it ranks second behind G-PCC v23 on a few point clouds such as \textit{Thaidancer} and \textit{Redandblack}.}
	

		{Table \ref{tab1} presents the quantitative BD-BR and BD-PSNR comparisons.} `Average' is the average value over 9 point clouds, while the `Average*' is the average result over all including point clouds marked with `*'.
	It can be observed that, excluding the point clouds marked with `*', the proposed TSC-PCAC achieves bit-rate savings ranging from 30.12\% to 50.14\%, and 36.62\% on average compared to the Sparse-PCAC. In terms of BD-PSNR, our method achieves BD-PSNR gain ranging from 1.15 dB to 2.35 dB, and 1.66 dB on average, which is significant.
	While including the seven point clouds marked with `*', {the proposed TSC-PCAC achieves bit rate savings ranging from 30.87\% to 54.00\%, and 38.53\% on average as compared with the Sparse-PCAC}. Meanwhile, the proposed TSC-PCAC achieves an average of 1.72 dB BD-PSNR gain.
	{Compared to NF-PCAC, TSC-PCAC achieves bitrate savings ranging from 12.85\% to 29.61\%, and 22.27\% on average. In terms of BD-PSNR, it achieves from 0.47 dB to 1.25 dB and 0.85 dB on average. While including point clouds marked with `*', the proposed TSC-PCAC achieves an average of 21.30\% bitrate saving and 0.80 dB BD-PSNR gain, which is significantly superior to the NF-PCAC.}

As compared to the G-PCC v23, the proposed TSC-PCAC saves bitrate up to 53.07\% across different point clouds. {However, as shown in Fig. \ref{fig5j} and Table \ref{tab1}, the proposed TSC-PCAC increases the BD-BR by 28.01\% for the \textit{Thaidancer}. This is mainly because the geometry and color attribute of \textit{Thaidancer} are relatively complex, and it has low PSNR value even at higher bit rates. These characteristics do not align with the training set, which primarily consists of point clouds with relatively simple color attributes.
Therefore, TSC-PCAC exhibits better compression performance for simpler-colored point clouds like \textit{basketball player} and \textit{dancer}. Increasing more colorful point clouds to training set may improve coding the complex sequences.} The TSC-PCAC is inferior to G-PCC v23 with bitrate increasing of 1.26\% on average. As we include the seven point clouds marked with `*', the proposed TSC-PCAC achieves an average of 11.19\% bit rate saving and 0.34 dB BD-PSNR gain, which proves the effectiveness of the proposed TSC-PCAC. {The TSC-PCAC achieves coding gains for most point clouds and is capable of learning from the first few frames to enhance coding the subsequent frames.}
		\begin{table*}[!t]
		\begin{center}
			\caption{Coding and decoding complexity comparison among the TSC-PCAC, G-PCC, Sparse-PCAC, NF-PCAC and Deep-PCAC. [Uint:s].}
			\label{tab3}
			\setlength{\tabcolsep}{1mm}
			\begin{tabular}{|c|cc|cc|cc|cc|cc|}
				\hline
				\multirow{2}{*}{PC} & \multicolumn{2}{c|}{Deep-PCAC}                       & \multicolumn{2}{c|}{Sparse-PCAC}                      & \multicolumn{2}{c|}{GPCC v23}                      & \multicolumn{2}{c|}{NF-PCAC}                       & \multicolumn{2}{c|}{TSC-PCAC}                      \\ \cline{2-11}
				& \multicolumn{1}{c|}{Enc. time}      & Dec. time      & \multicolumn{1}{c|}{Enc. time}      & Dec. time       & \multicolumn{1}{c|}{Enc. time}     & Dec. time     & \multicolumn{1}{c|}{Enc. time}     & Dec. time     & \multicolumn{1}{c|}{Enc. time}     & Dec. time     \\ \hline
				Soldier             & \multicolumn{1}{c|}{28.26}          & 23.77          & \multicolumn{1}{c|}{66.78}          & 294.57          & \multicolumn{1}{c|}{5.53}          & 4.95          & \multicolumn{1}{c|}{1.99}          & 2.45          & \multicolumn{1}{c|}{1.62}          & 4.49          \\ \hline
				Redandblack         & \multicolumn{1}{c|}{17.89}          & 14.88          & \multicolumn{1}{c|}{43.44}          & 151.94          & \multicolumn{1}{c|}{3.70}          & 3.31          & \multicolumn{1}{c|}{1.37}          & 1.67          & \multicolumn{1}{c|}{1.41}          & 3.39          \\ \hline
				Dancer              & \multicolumn{1}{c|}{80.02}          & 69.67          & \multicolumn{1}{c|}{156.90}         & 1135.91         & \multicolumn{1}{c|}{13.40}         & 12.08         & \multicolumn{1}{c|}{4.45}          & 5.48          & \multicolumn{1}{c|}{2.85}          & 9.79          \\ \hline
				Basketball   Player & \multicolumn{1}{c|}{91.21}          & 79.68          & \multicolumn{1}{c|}{173.35}         & 1364.30         & \multicolumn{1}{c|}{14.82}         & 13.37         & \multicolumn{1}{c|}{4.85}          & 5.98          & \multicolumn{1}{c|}{3.14}          & 10.66         \\ \hline
				Phil                & \multicolumn{1}{c|}{40.33}          & 33.99          & \multicolumn{1}{c|}{70.66}          & 323.55          & \multicolumn{1}{c|}{7.38}          & 6.62          & \multicolumn{1}{c|}{2.53}          & 3.09          & \multicolumn{1}{c|}{1.87}          & 5.54          \\ \hline
				Boxer               & \multicolumn{1}{c|}{25.41}          & 21.30          & \multicolumn{1}{c|}{60.92}          & 247.24          & \multicolumn{1}{c|}{5.04}          & 4.51          & \multicolumn{1}{c|}{1.74}          & 2.15          & \multicolumn{1}{c|}{1.46}          & 3.77          \\ \hline
				Thaidancer          & \multicolumn{1}{c|}{16.24}          & 13.42          & \multicolumn{1}{c|}{37.86}          & 122.97          & \multicolumn{1}{c|}{3.34}          & 2.99          & \multicolumn{1}{c|}{1.45}          & 1.76          & \multicolumn{1}{c|}{1.30}          & 3.22          \\ \hline
				Rafa                & \multicolumn{1}{c|}{21.84}          & 18.44          & \multicolumn{1}{c|}{48.19}          & 166.93          & \multicolumn{1}{c|}{4.30}          & 3.84          & \multicolumn{1}{c|}{1.63}          & 1.99          & \multicolumn{1}{c|}{1.55}          & 4.01          \\ \hline
				Sir Frederick       & \multicolumn{1}{c|}{24.58}          & 20.74          & \multicolumn{1}{c|}{53.60}          & 197.20          & \multicolumn{1}{c|}{4.76}          & 4.25          & \multicolumn{1}{c|}{1.61}          & 2.01          & \multicolumn{1}{c|}{1.50}          & 3.87          \\ \hline
				Andrew*             & \multicolumn{1}{c|}{34.54}          & 29.51          & \multicolumn{1}{c|}{58.95}          & 239.06          & \multicolumn{1}{c|}{6.53}          & 5.85          & \multicolumn{1}{c|}{2.33}          & 2.77          & \multicolumn{1}{c|}{1.65}          & 4.40          \\ \hline
				David*              & \multicolumn{1}{c|}{54.83}          & 46.92          & \multicolumn{1}{c|}{93.58}          & 473.10          & \multicolumn{1}{c|}{9.52}          & 8.55          & \multicolumn{1}{c|}{3.19}          & 3.86          & \multicolumn{1}{c|}{2.22}          & 7.41          \\ \hline
				Exerciese*          & \multicolumn{1}{c|}{75.59}          & 65.15          & \multicolumn{1}{c|}{148.08}         & 1062.44         & \multicolumn{1}{c|}{12.80}         & 11.52         & \multicolumn{1}{c|}{3.60}          & 4.69          & \multicolumn{1}{c|}{2.44}          & 7.85          \\ \hline
				Longdress*          & \multicolumn{1}{c|}{20.03}          & 16.60          & \multicolumn{1}{c|}{48.87}          & 178.55          & \multicolumn{1}{c|}{4.10}          & 3.67          & \multicolumn{1}{c|}{1.60}          & 1.94          & \multicolumn{1}{c|}{1.54}          & 4.17          \\ \hline
				Loot*               & \multicolumn{1}{c|}{19.91}          & 16.60          & \multicolumn{1}{c|}{47.90}          & 169.38          & \multicolumn{1}{c|}{4.03}          & 3.62          & \multicolumn{1}{c|}{1.46}          & 1.80          & \multicolumn{1}{c|}{1.34}          & 3.12          \\ \hline
				Model*              & \multicolumn{1}{c|}{71.52}          & 62.81          & \multicolumn{1}{c|}{141.21}         & 958.28          & \multicolumn{1}{c|}{12.40}         & 11.16         & \multicolumn{1}{c|}{4.05}          & 5.11          & \multicolumn{1}{c|}{2.90}          & 9.87          \\ \hline
				Sarah*              & \multicolumn{1}{c|}{36.40}          & 30.94          & \multicolumn{1}{c|}{63.85}          & 268.69          & \multicolumn{1}{c|}{6.74}          & 6.07          & \multicolumn{1}{c|}{1.97}          & 2.46          & \multicolumn{1}{c|}{1.59}          & 4.13          \\ \hline
				\textbf{Average*}    & \multicolumn{1}{c|}{\textbf{41.16}} & \textbf{35.28} & \multicolumn{1}{c|}{\textbf{82.13}} & \textbf{459.63} & \multicolumn{1}{c|}{\textbf{7.40}} & \textbf{6.65} & \multicolumn{1}{c|}{\textbf{2.49}} & \textbf{3.08} & \multicolumn{1}{c|}{\textbf{1.90}} & \textbf{5.61} \\ \hline
			\end{tabular}
		\end{center}
	\end{table*}

	\subsection{Visual Quality Evaluation}
	In addition, visual quality comparisons were performed to validate the effectiveness of TSC-PCAC.
 {
{Fig. \ref{fig6} illustrates the visual quality comparison of the compressed point clouds from five different encoding methods across two test sequences.} Figs. \ref{fig6a} to \ref{fig6g} display the enlarged images reconstructed by each encoding method for \textit{Redandblack}. To ensure fair comparisons, examples with similar bpp were selected. { The face of \textit{Redandblack} reconstructed from G-PCC v23, showed in Fig. \ref{fig6d}, exhibits noticeable block artifacts when compared to the original. The bitrate and PSNR for the reconstruction are 0.102 $bbp$ and 33.10 dB.} In contrast, bitrate and PSNR of the point clouds reconstructed by the Sparse-PCAC shown in Fig. \ref{fig6e} and NF-PCAC shown in Fig. \ref{fig6f} are 0.143 $bbp$ and 31.75 dB, and 0.116 $bbp$ and 32.21 dB, respectively. The reconstructed point clouds appear smoother with a lower PSNR. The reconstructed point cloud of our TSC-PCAC is presented in Fig. \ref{fig6g} with bitrate and PSNR of 0.132 $bpp$ and 33.24 dB. Compared with those from the Sparse-PCAC and NF-PCAC, it exhibits a lower bit rate and higher PSNR. Additionally, in terms of visual quality, it can be observed that TSC-PCAC is smoother and offers better visual quality. Similar results can be found for \textit{Soldier} from Figs. \ref{fig6so} to Fig. \ref{fig6m}. {The visual results further validate the superiority of TSC-PCAC for \textit{Redandblack} and \textit{Soldier} sequences comparing with Sparse-PCAC, NF-PCAC and G-PCC v23 methods.}
 }

\begin{figure*}[!t]
\centering
	\subfigure[] {
	\label{fig7a}
	\includegraphics[width=0.3\linewidth]{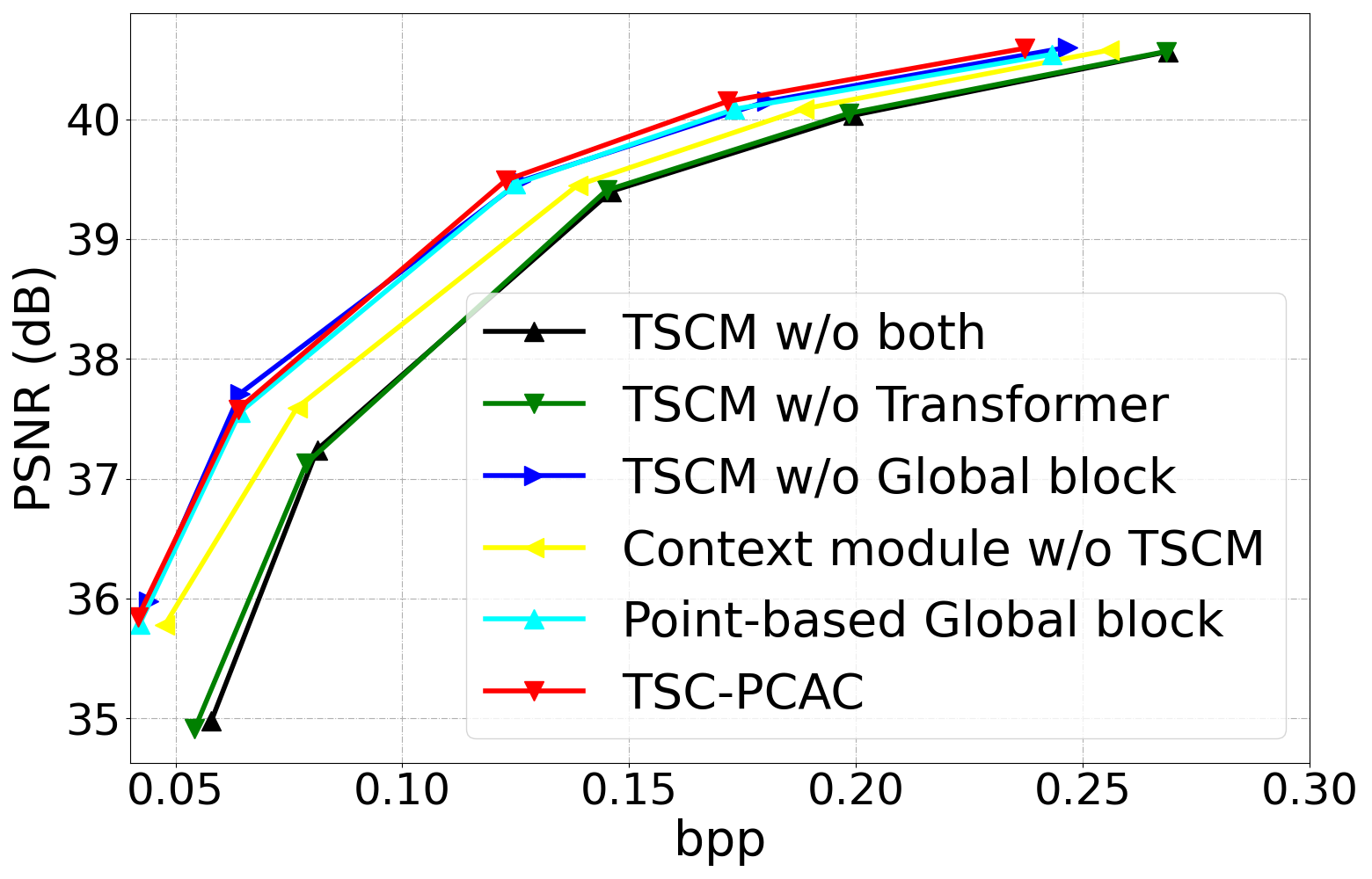}
	}
	\subfigure[] {
	\label{fig7b}
	\includegraphics[width=0.3\linewidth]{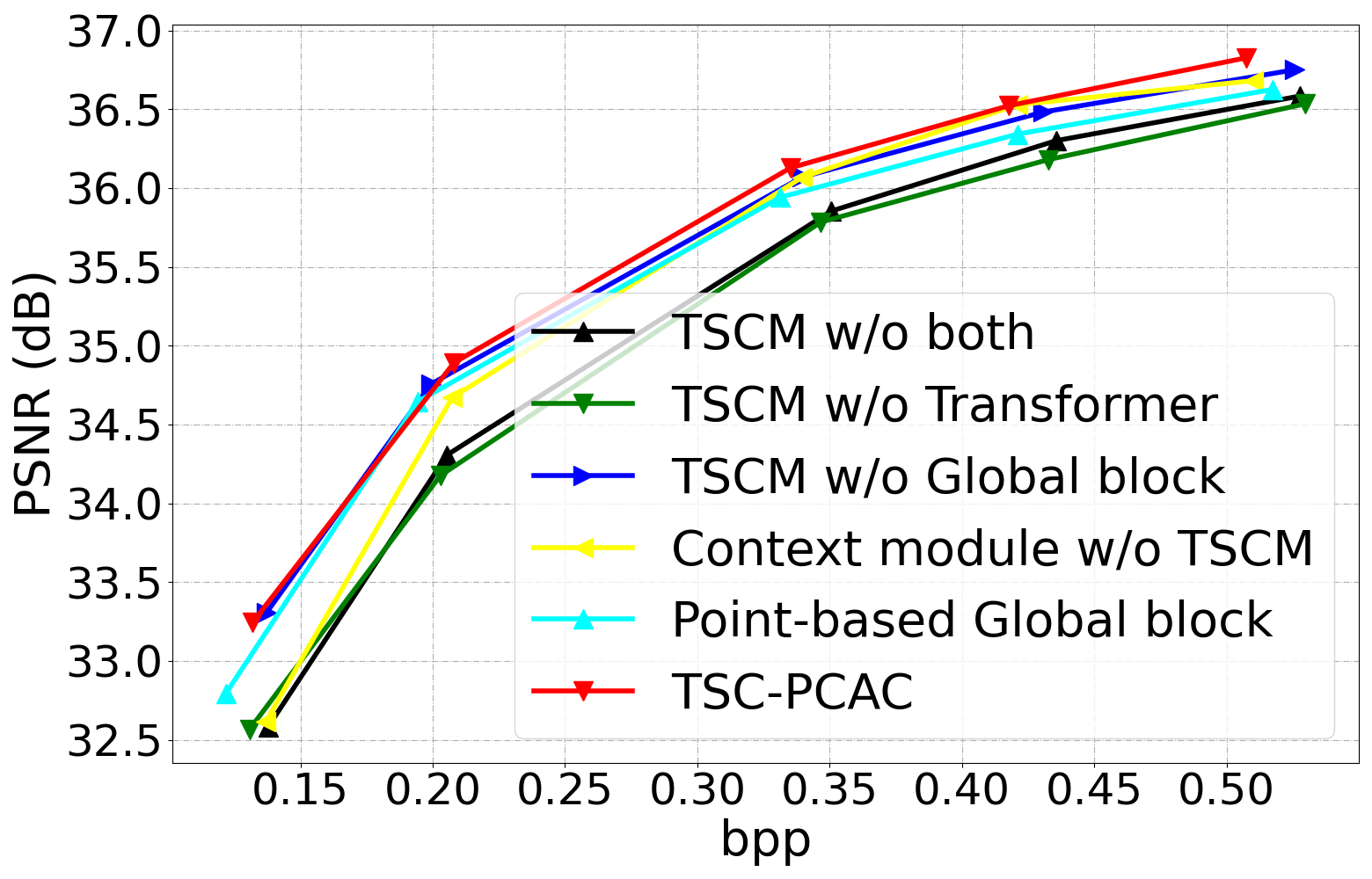}
	}
    \subfigure[] {
	\label{fig7c}
	\includegraphics[width=0.3\linewidth]{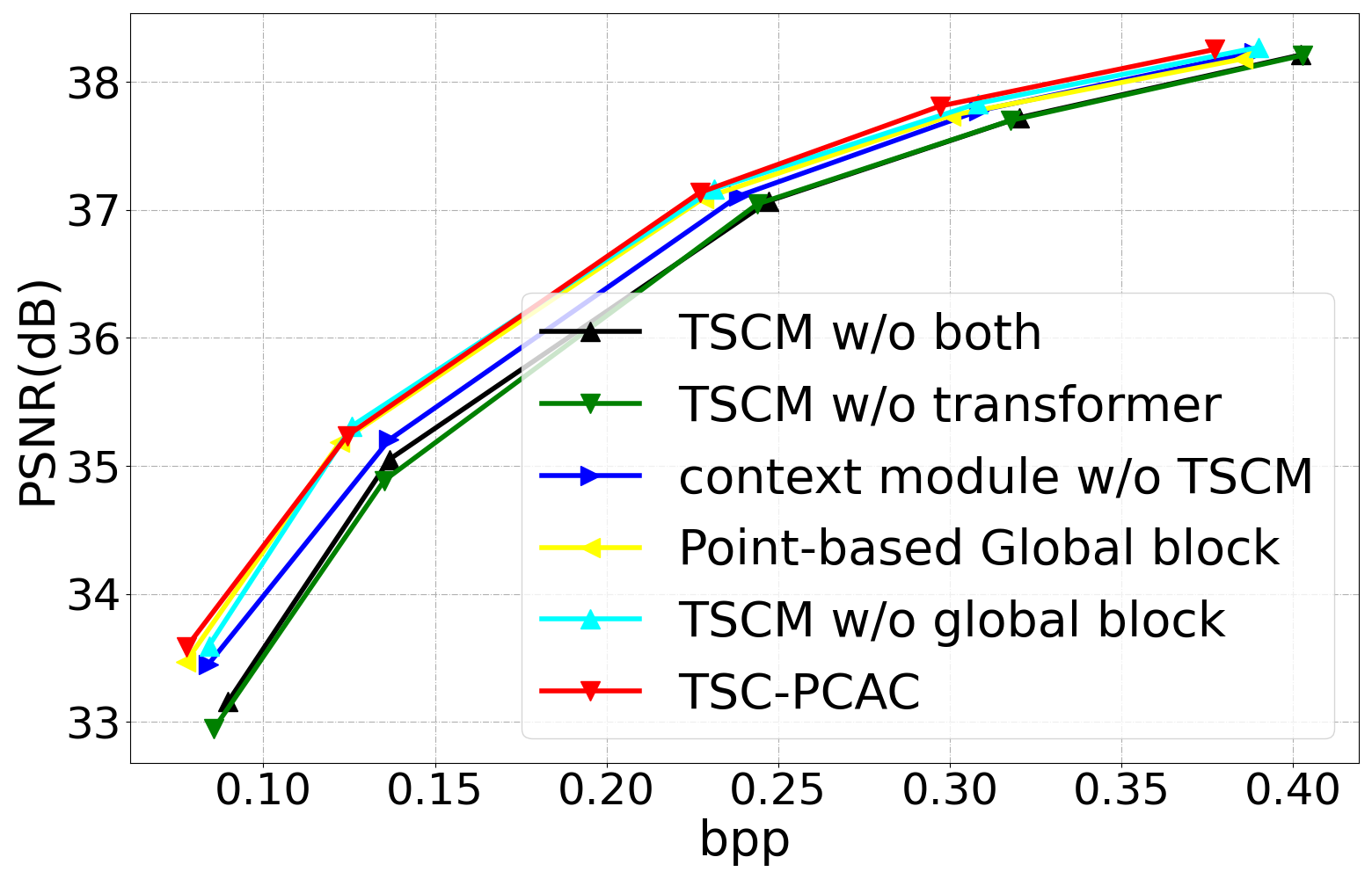}
	}
    \caption{Coding performance comparison for the TSC-PCAC ablation experiments. (a) Basketball Player, (b) Redandblack, (c) Average* of all sequences}
\label{fig7}
\end{figure*}
\subsection{Computational Complexity Evaluation}
	We also evaluated the computational complexity of TSC-PCAC on all the testing point clouds. We processed each point cloud sequentially using a single GPU and the operating system was Ubuntu 22.04.
	{For G-PCC, the encoding and decoding time for each point cloud corresponds to the average time required to encode and decode using different QP settings, while for the learned methods, it corresponds to the average time required to encode and decode using networks with different $\lambda_A$ values.}

\begin{table}[!t]
	\begin{center}
		\caption{{Computational complexity comparison}}
		\label{tab5}
		\setlength{\tabcolsep}{3mm}
		\begin{tabular}{|c|c|c|c|}
			\hline
			Methods        & Sparse-PCAC & NF-PCAC & TSC-PCAC \\ \hline
			Parameter      & 15.88M      & 51.16M  & 23.62M   \\ \hline
			Memory         & 4.75GB       & 16.69GB  & 5.58GB    \\ \hline
			Y Enc.         & 0.11s       & 1.10s   & 0.20s    \\ \hline
			Y Entropy Enc. & 82.02s      & 1.39s   & 1.70s    \\ \hline
			Total Enc.     & 82.13s      & 2.49s   & 1.90s    \\ \hline
			Y Entropy Dec. & 459.29s     & 1.13s   & 5.16s    \\ \hline
			Y Dec.         & 0.34s       & 1.95s   & 0.45s    \\ \hline
			Total Dec.     & 459.63s     & 3.08s   & 5.61s    \\ \hline
		\end{tabular}
	\end{center}
\end{table}
	From Table \ref{tab3}, it can be observed that Deep-PCAC and Sparse-PCAC have relatively long total encoding/decoding time, which are 41.16s/35.28s and 82.13s/459.63s, respectively. {The furthest point sampling in the network results in longer encoding and decoding time for Deep-PCAC. For Sparse-PCAC, the sequential processing of the voxel context module results in longer encoding and decoding time. This issue becomes particularly severe when processing denser point clouds, such as \textit{Basketball Player} and \textit{Dancer}, where the decoding time can exceed a thousand seconds.}
	 {For NF-PCAC, the average encoding/decoding time is 2.49s/3.08s}. For G-PCC v23, the encoding/decoding time ranges from 3.34s/2.99s to 14.82s/13.37s, with an average of 7.40s/6.65s. The encoding/decoding time of TSC-PCAC ranges from 1.30s/3.22s to 3.14s/10.66s, with an average of 1.90s/5.61s. TSC-PCAC reduces encoding/decoding time by 97.68\%/98.78\%, 23.74\%/-82.22\% and 74.33\%/15.66\% on average as compared with the Sparse-PCAC, NF-PCAC and G-PCC v23. Note that the encoding and decoding of G-PCC were performed on CPU, while {learning-based methods} were performed on the GPU+CPU platform.

{We also analyze the computational complexity for the modules of learned compression methods}. Table \ref{tab5} illustrates the network parameters, memory consumption in testing, the encoding and decoding time. {`Y Enc.' and `Y Dec.' represent the average computing time of encoder $E_{A}()$ and decoder $D_{A}()$.} Firstly, the parameter counts for Sparse-PCAC, TSC-PCAC, and NF-PCAC are 15.88M, 23.62M, and 51.16M respectively. NF-PCAC has the largest number of network parameters while Sparse-PCAC has the smallest. Secondly, the encoder and decoder runtime for Sparse-PCAC is the shortest, while NF-PCAC takes the longest. {However, the encoding/decoding time of Sparse-PCAC and TSC-PCAC are longer compared to NF-PCAC because of the context modules.} Sparse-PCAC employed the voxel context module performing autoregression in spatial dimensions, while our TSC-PCAC utilizes the channel context module performing autoregression in channel dimensions. Thus, TSC-PCAC greatly  reduces the number of autoregression steps. This results in longest entropy encoding/decoding time for Sparse-PCAC, moderate for TSC-PCAC, and shortest for NF-PCAC, which are 82.02s/459.29s, 1.70s/5.16s, and 1.39s/1.13s, respectively. Finally, the average total encoding time of the TSC-PCAC is 1.90s, which is the shortest among all schemes. The average total decoding time of the TSC-PCAC is 5.61s, which is in the second place and a little longer than that of the NF-PCAC, i.e., 3.08s.

\subsection{Ablation Study for TSC-PCAC}
	Ablation studies were performed to validate the effectiveness of the TSCM, the TSCM based channel context module, and the voxel-based global module. `TSCM w/o transformer' represents the transformer in TSCM is replaced with residual block, `TSCM w/o global block' represents the global block in TSCM is replaced with residual block, and `TSCM w/o both' represents the global block and transformer in TSCM are replaced with residual {block.} `Point-based Global block' indicates the TSCM that uses the global block from \cite{28}. 
		
{Table \ref{tab4} show the BD-BR and BD-PSNR gains achieved by the TSC-PCAC using different coding modules compared to the baseline Sparse-PCAC. Firstly, TSCM without Transformer achieves an average of 27.47\% bitrate saving and an average of 1.19 dB PSNR gain. If without both transformer and global block, the TSC-PCAC achieves an average bitrate saving of 28.38\% and 1.23 dB PSNR gain. As for the TSCM w/o Global block consisting of only residual blocks and the transformer, it achieves an average of 36.09\% bitrate savings. Secondly, to evaluate the effectiveness of using the TSCM, we replaced TSCM in the channel context module with two convolutional layers, denoted as `context module w/o TSCM'. The channel context module using conventional convolutions has larger parameters than that using TSCM. It is also found in Table \ref{tab4} that if removing TSCM from the channel context, only 32.16\% bitrate on average can be saved. Thirdly, to validate the effectiveness of using voxel-based global block in the TSC-PCAC, we replaced the voxel-based global block with point-based global block  \cite{28} in TSCM, and it is able to achieve an average of 36.89\% bit rate reduction. Finally, the overall TSC-PCAC achieves an average of 38.53\% bitrate saving, which is the best. Fig. \ref{fig7} shows the coding performance comparison of the ablation studies in TSC-PCAC, where the TSC-PCAC is the best for \textit{Basketball Player}, \textit{Redandblack} and average$*$ of all sequences. These results prove the effectiveness of each module in TSC-PCAC, including transformer and global block in TSCM, voxel-based global block and TSCM based channel context module.}
		
		\begin{table}[!t]
			\begin{center}
				\caption{Coding Performance Ablation Study for the TSC-PCAC.}
				\label{tab4}
				\setlength{\tabcolsep}{2.6mm}
				\begin{tabular}{|c|cc|}
					\hline
					\multirow{2}{*}{Methods}                       & \multicolumn{2}{c|}{Methods vs Sparse-PCAC}          \\ \cline{2-3}
					& \multicolumn{1}{c|}{BD-BR(\%)}     & BD-PSNR(dB) \\ \hline
					Context w/o TSCM                               & \multicolumn{1}{c|}{-32.16}          & 1.41          \\ \hline
					TSCM w/o Transformer                           & \multicolumn{1}{c|}{-27.47}          & 1.19          \\ \hline
					TSCM w/o both                                  & \multicolumn{1}{c|}{-28.38}          & 1.23          \\ \hline
					TSCM w/o Global block                          & \multicolumn{1}{c|}{-36.09}          & 1.60          \\ \hline
					\multicolumn{1}{|l|}{Point-based Global block} & \multicolumn{1}{c|}{-36.89}          & 1.62          \\ \hline
					TSC-PCAC                                       & \multicolumn{1}{c|}{\textbf{-38.53}} & \textbf{1.72} \\ \hline
				\end{tabular}
			\end{center}
		\end{table}

	\section{Conclusions}
	\label{section4}
	In this paper, we propose a voxel Transformer and Sparse Convolution-based Point Cloud Attribute Compression (TSC-PCAC) for point cloud broadcasting, where an efficient Transformer and Sparse Convolution Module~(TSCM) and TSCM based channel context module are developed to improve the attribute coding efficiency. The TSCM integrates local dependencies and global spatial-channel features for point clouds, modeling local correlations among voxels and {capturing} global features to remove redundancy. Furthermore, the TSCM based channel context module exploits inter-channel correlations to improve the predicted probability distribution of quantized latent representations. The TSC-PCAC achieves an average of 38.53\%, 21.30\%, and 11.19\% bit rate reductions as compared with the Sparse-PCAC, NF-PCAC and G-PCC v23. Meanwhile, it maintains low encoding/decoding complexity. Overall, the proposed TSC-PCAC is effective for point cloud attribute coding.
	
	{The deep learning based Sparse-PCAC, NF-PCAC, and TSC-PCAC are for attribute coding only and require to be trained multiple times for different rate points. In future, we will investigate deep networks for higher compression ratio, as well as joint geometry and attribute coding.}
	
	
	\bibliographystyle{IEEEtran}
	\bibliography{IEEEabrv,aaaaa}
	
\end{document}